\documentclass[aps,prxquantum,reprint,twocolumn,superscriptaddress,floatfix,nofootinbib,longbibliography]{revtex4-1}
\usepackage{amsmath,amssymb,color,comment,physics,mathtools}
\usepackage[makeroom]{cancel}
\usepackage[caption=false]{subfig}
\usepackage{mathrsfs}
\usepackage{graphicx}
\usepackage{subfig}
\usepackage[countmax]{subfloat}
\usepackage[english]{babel}
\usepackage[bookmarks=true,colorlinks,linkcolor=OrangeRed,urlcolor=NavyBlue,citecolor=RoyalBlue]{hyperref}
\usepackage[dvipsnames]{xcolor}
\usepackage{ulem} 
\usepackage{braket}
 \usepackage{dsfont}

\definecolor{mygreen}{rgb}{0,0.5,0}
\definecolor{myblue}{rgb}{0,0,0.75}
\definecolor{mymagenta}{cmyk}{0,1,0,0.12}
\definecolor{mygray}{rgb}{0.5,0.5,0.5}

\def\d{\mathrm d}

\definecolor{mypink1}{rgb}{0.858, 0.188, 0.478}
\definecolor{mypurple}{rgb}{0.49,0.18,0.56}
\definecolor{mygold}{rgb}{0.93,0.69,0.13}
\definecolor{mygreen}{rgb}{0,0.5,0}
\definecolor{myred}{rgb}{1,0,0}
\definecolor{myblue}{rgb}{0,0,0.75}
\definecolor{mymagenta}{cmyk}{0,1,0,0.12}
\definecolor{mygray}{rgb}{0.5,0.5,0.5}

\newcommand{\ignore}[1]{}
\usepackage{geometry}

\begin{document}
\title{Gauge-Symmetry Protection Using Single-Body Terms}

\author{Jad C.~Halimeh}
\affiliation{INO-CNR BEC Center and Department of Physics, University of Trento, Via Sommarive 14, I-38123 Trento, Italy}

\author{Haifeng Lang}
\affiliation{INO-CNR BEC Center and Department of Physics, University of Trento, Via Sommarive 14, I-38123 Trento, Italy}
\affiliation{Theoretical Chemistry, Institute of Physical Chemistry, Heidelberg University, Im Neuenheimer Feld 229, 69120 Heidelberg, Germany }

\author{Julius Mildenberger}
\affiliation{INO-CNR BEC Center and Department of Physics, University of Trento, Via Sommarive 14, I-38123 Trento, Italy}

\author{Zhang Jiang}
\affiliation{Google AI Quantum, Venice, CA, USA}

\author{Philipp Hauke}
\affiliation{INO-CNR BEC Center and Department of Physics, University of Trento, Via Sommarive 14, I-38123 Trento, Italy}

\begin{abstract}
Quantum-simulator hardware promises new insights into problems from particle and nuclear physics. A major challenge is to reproduce gauge invariance, as violations of this quintessential property of lattice gauge theories can have dramatic consequences, e.g., the generation of a photon mass in quantum electrodynamics. Here, we introduce an experimentally friendly method to  protect gauge invariance in $\mathrm{U}(1)$ lattice gauge theories against coherent errors in a controllable way. Our method employs only single-body energy-penalty terms, thus enabling practical implementations. As we derive analytically, some sets of penalty coefficients render undesired gauge sectors inaccessible by unitary dynamics for exponentially long times, and, for few-body error terms, with resources independent of system size. These findings constitute an exponential improvement over previously known results from energy-gap protection or perturbative treatments. In our method, the gauge-invariant subspace is protected by an emergent global symmetry, meaning it can be immediately applied to other symmetries. In our numerical benchmarks for continuous-time and digital quantum simulations, gauge protection holds for all calculated evolution times (up to $t>10^{10}/J$ for continuous time, with $J$ the relevant energy scale). Crucially, our gauge-protection technique is simpler to realize than the associated ideal gauge theory, and can thus be readily implemented in current ultracold-atom analog simulators as well as digital noisy intermediate scale quantum (NISQ) devices.
\end{abstract}

\date{\today}
\maketitle

\section{Introduction}
Quantum simulation promises to solve complex quantum many-body systems using dedicated quantum hardware. A particularly appealing target for quantum simulation is the solution of lattice gauge theories (LGTs) \cite{Wiese_review,Dalmonte_review,Pasquans_review}. Thanks to their fundamental importance in high-energy and nuclear physics, gauge theories are currently one of the main drivers of developments in scientific high-performance computing \cite{Joo2019,Nagel_book}, 
and they have deep connections to topological phases of matter and topological quantum computing \cite{Nussinov2009,Wen2013}.
The goal of quantum simulation is to advance into the regime of a true quantum advantage \cite{Arute2019}, i.e., a regime that is no longer accessible even for classical supercomputers, e.g., large many-body systems far from equilibrium. As long as fully fault-tolerant quantum computers are still out of reach, it is crucial to design feasible error-mitigation strategies that ensure the reliable working of the current noisy quantum simulators \cite{Hauke_review}.  

\begin{figure}[!ht]
	\centering
	\includegraphics[width=.48\textwidth]{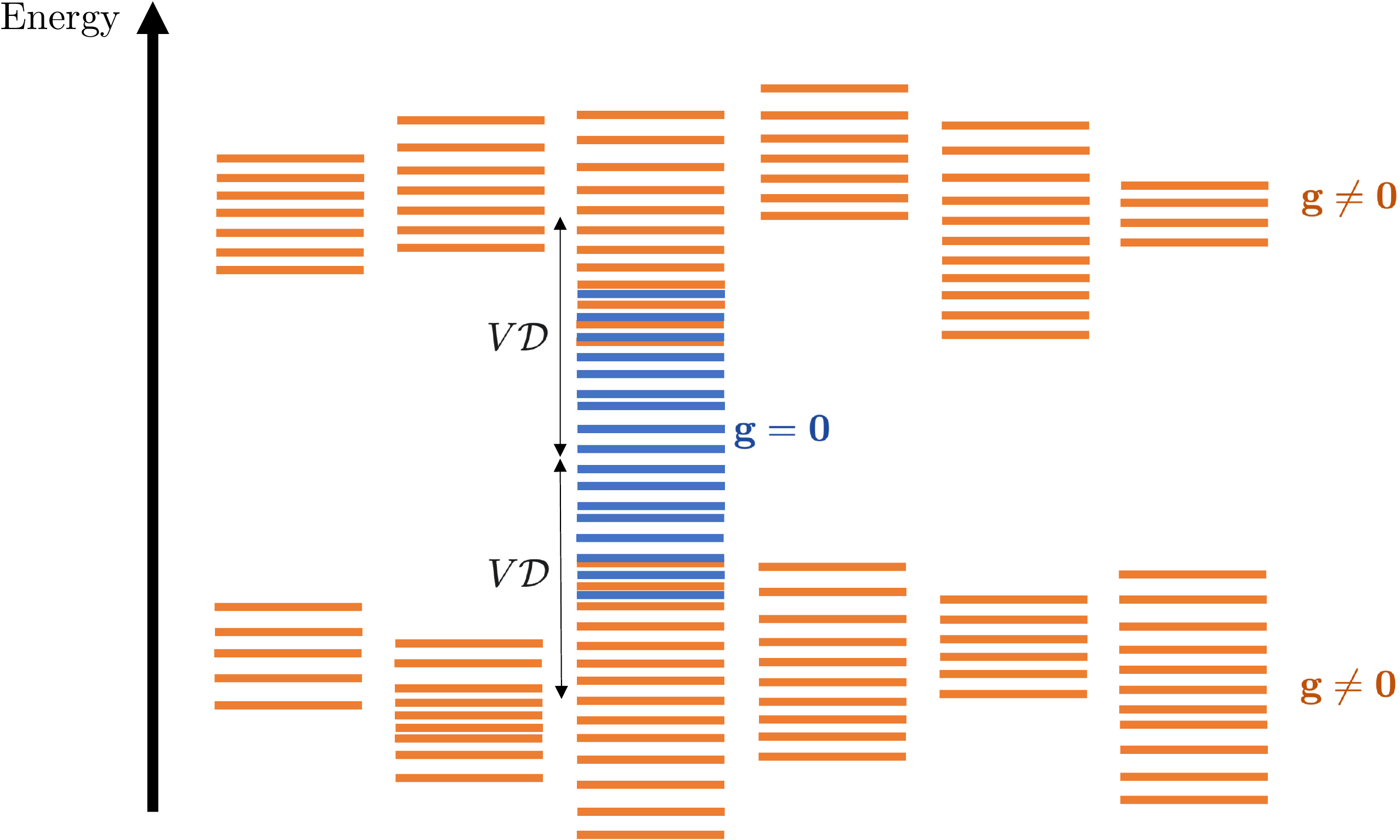}
	\caption{(Color online). A lattice gauge theory can be decomposed into symmetry sectors characterized by an extensive number of local conserved quantities, defined by the gauge-symmetry generators $G_j$ with eigenvalues $\mathbf{g}=(\ldots,g_j,\ldots)$. 
	In realistic quantum simulator experiments, coherent error terms $\lambda H_1$ can break this gauge symmetry. As we prove analytically, reliable gauge invariance can nevertheless be dynamically achieved by introducing the gauge protection $VH_G=V\sum_jc_jG_j$, which is composed of single-body terms proportional to the gauge-symmetry generators, weighted by appropriate coefficients $c_j$. 
	When the coefficients satisfy Eq.~\eqref{eq:c_j}, the protection term shifts undesired sectors by an energy scale $\mathcal{D}=\min_{\mathbf{g}\neq\mathbf{0}}|\mathbf{c}^\intercal\cdot\mathbf{g}|$. At sufficiently large $V$, $VH_G$ induces an emergent global symmetry that in the sector $\mathbf{g}=\mathbf{0}$ coincides with the local gauge invariance. As a consequence, these experimentally simple single-body terms suppress gauge violations $\sim(\lambda/V)^2$ for exponentially long times and---for local error terms---independently of system size, thereby bringing the dynamics perturbatively close to a renormalized ideal gauge theory.}
	\label{fig:illustration}
\end{figure}

In the case of a gauge theory, this reliability is particularly delicate: the associated gauge and matter fields necessarily have to obey a precise local conservation law known as a gauge symmetry---an example is Gauss's law in quantum electrodynamics (QED). Quantum-simulator realizations that do not enjoy a natural law imposing gauge symmetry \cite{Goerg2019,Schweizer2019,Mil2019,Yang2020} will always suffer from microscopic terms that coherently break gauge invariance. Violations of gauge invariance can have dramatic consequences: for example, they can generate a photon mass in QED, which would reduce the infinite-range Coulomb law to a Yukawa potential. In equilibrium, a massless photon can still emerge in a renormalized theory if the gauge-breaking terms are sufficiently small \cite{Foerster1980,Poppitz2008}. Out of equilibrium, recent works have found indications that gauge invariance presents a certain intrinsic robustness as gauge errors in intermediate-scale quantum devices can build up slowly \cite{Halimeh2020b,Halimeh2020c}. Yet, it remains an outstanding challenge to devise schemes that enable an active, controllable protection of gauge invariance and which are implementable in realistic quantum hardware.

Here, we present a scheme that uses simple single-body terms to controllably protect nonequilibrium dynamics against unitary breaking of gauge invariance, at least up to exponentially long times and---for local error terms---independent of system size. 
The scheme is based on adding an energy penalty consisting of a linear sum of the local generators of Gauss's law, with weights chosen according to an equation that we derive. 
As a result, a global symmetry is enforced that within the target gauge-invariant sector acts as the desired local gauge symmetry. 
We analytically prove the gauge protection by adapting results from periodically driven systems \cite{Abanin2017}, thus providing a firm theoretical framework for our protection scheme. 
These results can immediately be extended to improve analytical predictions for other scenarios, such as energy-gap protection using stabilizer codes \cite{Jordan2006,Young2013}.
Moreover, using results from the quantum Zeno effect in coherent systems \cite{facchi2002quantum,facchi2004unification,facchi2009quantum,burgarth2019generalized}, we show that weaker (and potentially experimentally even simpler) forms of penalties protect gauge invariance at least to polynomially long times. As a corollary, we prove the same type of protective strength for the previously proposed, experimentally more challenging two-body protection terms.
Further, using exact numerics, we illustrate the scheme for an Abelian $\mathrm{U}(1)$ gauge theory in one spatial dimension and demonstrate the controllable protection for all simulated times. 
The proposed single-body protection is considerably more experimentally friendly than previous proposals, which rely on engineered spatially correlated noise \cite{Stannigel2014,Lamm2020}, on energy penalties that require precisely tuned two-body interactions \cite{Banerjee2012,Hauke2013,Kuehn2014,Kuno2015,Yang2016,Kuno2017,Negretti2017,Barros2019,Halimeh2020a}, or multi-qubit operations \cite{Young2013}. 
As such, the protection scheme can be implemented in digital devices using single-qubit gates \cite{Arute2019} as well as in state-of-the-art analog realizations, e.g., in optical lattices using a site-dependent chemical potential \cite{Mil2019,Yang2020}. 
Thus, our work opens a pathway for controlled gauge invariance in large-scale LGT quantum simulators. Even more, our results can be extended immediately to scenarios with global symmetries. 

Our paper is organized as follows: We start with some background on the considered gauge theory and protection scheme in Sec.~\ref{sec:background}. In Sec.~\ref{sec:theorem}, we rigorously derive the \textit{gauge protection theorem}, which underlies the basis of our work. Using this theorem, in Sec.~\ref{sec:results} we demonstrate the gauge protection numerically through exact diagonalization calculations of gauge-violation dynamics in the $\mathrm{U}(1)$ gauge theory simulating an analog quantum simulator and a digital quantum computer. In Sec.~\ref{sec:discussion}, we outline the connection of our findings to dynamical decoupling and energy gap protection as well as their application to ongoing cold-atom experiments. We conclude in Sec.~\ref{sec:conclusion}. Several Appendices complement the results and discussions of the main text.

\section{Background}\label{sec:background}

Consider a quantum-simulation experiment aiming at implementing a $(d+1)$-dimensional lattice gauge theory (i.e., $d$ spatial dimensions). The theory is described by an ideal Hamiltonian $H_0$ that is invariant under gauge transformations with the unitaries $e^{-i\alpha_j G_j}$, generated by local Gauss's-law generators $G_j$, where $j$ are matter sites of the gauge theory. That is, $[H_0,G_j]=0$, $\forall j$. 
In this study, we focus on Abelian LGTs, i.e., $[G_j,G_l]=0$.
Further, we denote the (integer) spectrum of $G_j$ as $\{g_j\}$ and define $\textbf{g}$ as the vector of $g_j$. 
Without restriction of generality, we choose the target gauge-invariant subspace as consisting of those states that fulfill $G_j\ket{\psi}=0$, $\forall j$, i.e., the sector $\textbf{g}=\textbf{0}$ (also called `physical subspace'). 	

Though our discussion holds more generally, we will illustrate it numerically in Sec.~\ref{sec:results} using a quantum link model (QLM) in 1+1 dimensions, given by the Hamiltonian \cite{Wiese_review,Chandrasekharan1997} 
\begin{align}\label{eq:H0}
H_0 = \sum_{j=1}^L\Big[J\big(\sigma^-_j\tau^+_{j,j+1}\sigma^-_{j+1}+\text{H.c.}\big)+\frac{\mu}{2}\sigma^z_j\Big].
\end{align}
Here, the Pauli ladder operators $\sigma^\pm_j$ ($\tau^\pm_{j,j+1}$) are the creation/annihilation (flipping) operators of the matter (gauge) field at site $j$ [link $(j,j+1)$]. Accordingly, $\sigma^z_j$ ($\tau^z_j$) is a mass-density (electric-field) operator at site $j$ [link $(j,j+1)$]. The matter--gauge coupling strength is given by $J$, the matter rest mass is $\mu$, and the number of matter sites is $L$, where periodic boundary conditions are assumed, meaning that there are also $L$ links in the model. 
The local-symmetry generators of the $\mathrm{U}(1)$ QLM in Eq.~\eqref{eq:H0} are 
\begin{align}\label{eq:Gj}
G_j=\frac{(-1)^j}{2}\big(\sigma^z_j+\tau^z_{j-1,j}+\tau^z_{j,j+1}+1\big).
\end{align}
(See Appendix~\ref{sec:appMoreAboutU1QLM} for details on the eigenvalues of $G_j$ in this model.)
For concreteness, it may be instructive to have Eqs.~\eqref{eq:H0} and \eqref{eq:Gj} in mind, but the following considerations hold for arbitrary LGTs that satisfy the commutation relations $[G_j,G_l]=[H_0,G_j]=0$, $\forall j,l$. 
In an idealized quantum simulation that would perfectly implement such a $H_0$, the $G_j$ would be conserved quantities.

In realistic implementations without fine tuning (and not using certain encoding strategies \cite{Martinez2016,Muschik2017,Bernien2017,Surace2019,Zohar2019}), however, there will be coherent terms that break gauge invariance, which we subsume in the error term $\lambda H_1$, with $[H_1,G_j]\neq 0$; $\lambda$ controls the error strength, which in realizations such as in Refs.~\cite{Schweizer2019,Mil2019,Yang2020} may be small but nonnegligible. These terms will drive the quantum simulator out of the target gauge-invariant subspace. 

Several proposals have been made regarding how to protect against such gauge-invariance breaking using energy-penalty terms quadratic in the Gauss's law generators  \cite{Zohar2011,Zohar2012,Banerjee2012,Zohar2013,Hauke2013,Kuehn2014,Kuno2015,Yang2016,Kuno2017,Negretti2017,Barros2019,Halimeh2020a}, 
\begin{align}\label{eq:FullProtection}
V \tilde{H}_G=V\sum_j G_j^2,
\end{align}
with protection strength $V>0$. 
Such a term shifts all states not in the target sector $\textbf{g}=\textbf{0}$ up in energy, such that the desired physics can be reproduced in a controlled manner. 
Indeed, such a scheme has been shown to give rise to two distinct regimes in an out-of-equilibrium simulation starting from a gauge-invariant initial state \cite{Halimeh2020a}: an uncontrolled-violation regime when $V/\lambda$ is small, and a controlled-error regime in the case of large enough $V/\lambda$. In the controlled regime, the system is shown in degenerate perturbation theory to be perturbatively close to a renormalized ideal gauge theory. 
The implementation of a penalty quadratic in the Gauss's-law generators as in Eq.~\eqref{eq:FullProtection}, however, poses formidable experimental challenges, as it requires the precise design of two-body interaction terms involving the matter and gauge fields. 

As the main result of our work, in the next section we will show analytically that a protection linear in the Gauss's-law generators---and thus comprised of only single-body terms---suffices to ensure a controlled violation in the gauge-theory dynamics:
\begin{align}\label{eq:LinearProtection}
V H_G=V\sum_j c_j G_j.
\end{align}
The full Hamiltonian describing the envisioned quantum-simulator experiment is then given by 
\begin{equation}
H = H_0+\lambda H_1+VH_G. 
\label{equ:effectiveHamiltonian}
\end{equation}
To achieve gauge-protected dynamics, it is important that the system be prepared in an initial state that resides in the target sector $\textbf{g}=\textbf{0}$. Even if there exist undesired gauge sectors at higher and lower energy, the system remains dynamically constrained to the target sector (see Fig.~\ref{fig:illustration}). This is in contrast to Eq.~\eqref{eq:FullProtection}, where the target sector $\textbf{g}=\textbf{0}$ becomes lowest in energy, and which thus allows not only for protected dynamics but also for a controlled cooling into the ground state. However, the present scenario is in line with ongoing cold-atom experiments, which currently either consider quench dynamics \cite{Goerg2019,Schweizer2019,Mil2019} or adiabatic transfer across phase transitions \cite{Yang2020}, in both cases starting from simple states within the target sector $\textbf{g}=\textbf{0}$. 

In a worst-case scenario, the coefficients $c_j$ in Eq.~\eqref{eq:LinearProtection} have to comply with Eq.~\eqref{eq:c_j} derived in the next section, such that resonances between gauge violations at different sites are avoided regardless of the form of gauge invariance-breaking terms in $H_1$. 
However, such a `compliant' sequence is only necessary in the case of an extreme error (such as $H_1$ given in Eq.~\eqref{eq:H1} below). As we shall illustrate, in more benign situations, such as local errors (see Fig.~\ref{fig:AnalogDynamics_experiment}) or desired protection only up to times polynomial in $V$ (see Appendix~\ref{sec:QZE}), modified sequences of $c_j$ can suffice, meaning there is room for inaccuracies in the implementation. 

Notably, the type of protection given in Eq.~\eqref{eq:LinearProtection} can be experimentally realized using \textit{single-body} terms. As we will demonstrate in Secs.~\ref{sec:digitalCircuit} and~\ref{sec:coldatom}, these can be simple single-qubit gates in digital circuits, respectively single-site chemical potentials in optical-lattice implementations. 
The proposed realization is thus not only considerably less challenging than the two-body terms of Eq.~\eqref{eq:FullProtection}, it is also advantageous with respect to other proposals based on engineered noise. For example, according to the scheme of Ref.~\cite{Stannigel2014} classical dephasing should be added that is correlated between a matter site $j$ and its neighboring links, such that it couples to the Gauss's-law generator $G_j$, but without correlations across matter sites. 
Such a noise can suppress coherent gauge breaking so it induces only a slow diffusion out of the target subspace and gauge violations occur on times polynomially large in the noise strength. In contrast, as we will demonstrate now, our coherent scheme limits the leakage out of the gauge-invariant subspace to a controlled and perturbatively small value, at least up to times exponentially large in $V$. 

\section{Gauge protection theorem}\label{sec:theorem}

To demonstrate the gauge protection, we adapt results from Ref.~\cite{Abanin2017} on slow heating in periodically driven systems. The aim is to transform the full Hamiltonian of Eq.~\eqref{equ:effectiveHamiltonian} into a theory perturbatively close to a renormalized version of the original one that manifests as an approximate preservation of $H_G$ once the associated energy scale $V$ dominates. 
Define $\Pi_n$ the projection operator onto eigenstates of $H_G$ with eigenvalue $n$.
Then, $H$ can be decomposed into an $H_G$-invariant part, $H_{\rm diag}+V H_G$ with  $H_{\rm diag}\coloneqq\sum_n \Pi_n (H_0+\lambda H_1)\Pi_n = H_0+\lambda \sum_n \Pi_n H_1\Pi_n$, and the remainder, $H_{\rm ndiag}=H-H_{\rm diag}-VH_G$. By construction, $[H_{\rm ndiag},H_G]\neq 0$ and $[H_{\rm diag},H_G]=0$ (though in general $[H_{\rm diag},G_j]\neq 0$; i.e., $H_{\rm diag}$ obeys a global symmetry generated by $H_G$, but not the local gauge symmetry generated by $G_j$; we will come back to this point further below).

Before proceeding, it is convenient to introduce the algebra and a family of norms as follows \cite{Abanin2017}. Being interested in a lattice gauge theory on a cubic lattice in $d$ spatial dimensions, we define $\Lambda$ as a finite subset of the lattice $\mathbb{Z}^d$. Define $\mathcal{B}_\Lambda$ as the algebra of bounded operators acting on the total Hilbert Space $\mathcal{H}_\Lambda$, equipped with the standard operator norm. We also define the subalgebra $\mathcal{B}_S \subset \mathcal{B}_\Lambda$ of operators of the form $O_S\otimes\mathcal{I}_{\Lambda \backslash S}$ with $S\subset \Lambda$. 
Any operator $X$ can be decomposed (in a nonunique way) as $X = \sum_{S\in \mathcal{P}_c(\Lambda)} X_S$ where $X_S \in \mathcal{B}_S$ and $\mathcal{P}_c(\Lambda)$ denotes the set of finite, connected (by adjacency) subsets of $\Lambda$. The collection $X_S$ is referred to as an (interaction) potential. Define a family of norms on potentials, parametrized by a rate $\kappa >0$ that gives different weights to operators with different spatial support, 
\begin{equation}
||X||_\kappa \coloneqq \sup\limits_{x\in\Lambda} \sum_{S\in \mathcal{P}_c(\Lambda):S\ni x} e^{\kappa |S|} ||X_S||.
\end{equation}
The supremum in this definition chooses the lattice site $x$ with the largest sum of weighted norms of the operators $X_S$ that have support on $x$.

Assume there exists a $\kappa_0$ such that the energy scale can be defined as $
V_0 \coloneqq \frac{54\pi}{\kappa_0^2}(||H_{\rm diag}||_{\kappa_0}+2||H_{\rm ndiag}||_{\kappa_0})
$. Theorem 3.1 from Ref.~\cite{Abanin2017} then states the following: If the spectrum of $c_jG_j$ are integers for all $j$ and $V$ fulfils the conditions
\begin{equation}
\label{eq:Vge}
V \ge \frac{9\pi||H_{\rm ndiag}||_{\kappa_0}}{\kappa_0}\,
\end{equation}
and 
\begin{equation}
\label{eq:nstar}
n_* \coloneqq \lfloor \frac{V/V_0}{(1+\ln{V/V_0})^3}\rfloor - 2 \ge 1,
\end{equation}
there exists a unitary operator $Y$ such that 
	\begin{align}\nonumber
	YHY^\dag&=VH_G + H^\prime \\
	&= VH_G + H^\prime_{\rm diag} + H^\prime_{\rm ndiag},
	\end{align}
	with $H^\prime = YHY^\dag - VH_G$, $H^\prime_{\rm diag}=\sum_n \Pi_n H^\prime\Pi_n$, $H^\prime_{\rm ndiag}=H^\prime-H^\prime_{\rm diag}$, 
	and
	\begin{align}
	||H^\prime_{\rm diag}-H_{\rm diag}||_{\kappa_{n_*}}&\le C(V_0/V), \\
	||H^\prime_{\rm ndiag}||_{\kappa_{n_*}}&\le (2/3)^{n_*}||H_{\rm ndiag}||_{\kappa_0},
	\end{align}
	where $\kappa_{n_*}\coloneqq\kappa_0[1+\log(1+n_*)]^{-1}$ and $C$ is a constant. 
	In other words, $H^\prime_{\rm diag}$ is perturbatively close (in $V_0/V$) to $H_{\rm diag}$ and the new contribution that fails to commute with $H_G$, $H^\prime_{\rm ndiag}$, is exponentially small (in $V/V_0$).  
	
Furthermore, $Y$ is quasilocal and close to the identity in the sense that for any local operator $X$,
	\begin{equation}
	||YXY^\dag-X||_{\kappa_{n_*}}\le C(V_0/V)||X||_{\kappa_0}.
	\end{equation} 
Then, for arbitrary local operator $O$ and up to an exponentially large time $t$ on the scale $e^{kn_*}/V_0$, we have 
\begin{align}
\label{eq:norm_of_operator}
&||U(t)^\dag OU(t)-e^{it(VH_G+H^\prime_{\rm diag})}Oe^{-it(VH_G+H^\prime_{\rm diag})}||\nonumber
\\ 
&\le \frac{K(O)}{V},
\end{align}
where $U(t)=e^{-iHt}$ is the full time-evolution operator, $0<k<\frac{1}{d+1}\ln{(3/2)}$, and $K(O)$ is $V$- and volume-independent. 

Until now, $H_G$ is a general operator. As we have seen, for sufficiently large $V$ it defines an emergent \textit{global} symmetry that the full Hamiltonian $H$ preserves up to exponentially long times. This scheme can thus be used to protect an arbitrary global symmetry with integer spectrum \cite{Abanin2017}. We can use that to derive an \textit{effective} preservation of a \textit{local} symmetry. 
To this end, we now specialize to our case of $H_G=\sum_j c_j G_j$. 
 
In a typical quench experiment, the initial state $\ket{\psi_0}$ is prepared in the gauge sector $\textbf{g}=\textbf{0}$. We can quantify the leakage out of this target gauge sector by substituting $O=G_j$ in Eq.~\eqref{eq:norm_of_operator} and using $ ||U(t)^\dag  G_jU(t)-e^{itH^\prime_{\rm diag}}G_je^{-itH^\prime_{\rm diag}}||\leq  K(G_j)/V$ to estimate $|\braket{U(t)^\dag  G_jU(t)}-\braket{e^{itH^\prime_{\rm diag}}G_je^{-itH^\prime_{\rm diag}}}|\leq K(G_j)/V$. In general, $[H_G,G_j]=[H_G,H^\prime_{\rm diag}]=0$, but $[G_j,H^\prime_{\rm diag}]\neq 0$. We can nevertheless achieve the protection of $|\braket{U(t)^\dag  G_jU(t)}|$ if the projector onto the zero eigenvalue of $H_G$, $\Pi_{n=0}$, projects only onto the states with $\textbf{g}=\textbf{0}$. Then, $e^{-iH^\prime_{\rm diag}t}\ket{\psi_0}$ remains in the gauge sector $\textbf{g}=\textbf{0}$, which yields $\braket{e^{itH^\prime_{\rm diag}}G_je^{-itH^\prime_{\rm diag}}}=0$. 
Then, the gauge violation remains bounded by a perturbatively small value, $|\braket{U(t)^\dag  G_jU(t)}|\leq K(G_j)/V$, up to exponentially long times $t\sim \mathcal{O}(\frac{1}{V_0}e^{V/V_0})$, and---for error terms with finite bounded support---independent of system size. 
In this way, we have designed a global symmetry operator $H_G$ such that within the sector $\textbf{g}=\textbf{0}$ it approximates the local gauge invariance with certified error.  

One way to satisfy the condition of the zero eigenvalue of $H_G$ coinciding with $\textbf{g}=\textbf{0}$ is by designing the integer coefficients $c_j$ such that  
\begin{equation}
	\label{eq:c_j}
	\sum_j c_jg_j=0 \quad \mathrm{iff} \quad \textbf{g}=\textbf{0}.
\end{equation}
In what follows, we refer to sequences that fulfil this condition as \textit{compliant}. 
Moreover, below we normalize to set the $c_j$ with maximal absolute value to unity, such that $V$ encodes the overall scale of the gauge-protection term.

Using a compliant sequence ensures that resonances between gauge violations at different sites are avoided regardless of the form of gauge invariance-breaking terms.
Importantly, the resulting gauge protection can take place even if the different gauge sectors are not energetically well separated. 
The gauge sector closest to the target sector $\mathbf{g}=\mathbf{0}$ lies at an energy $V\mathcal{D}=\min_{\mathbf{g}\neq \mathbf{0}}V|\mathbf{c}^\intercal\cdot\mathbf{g}|$ (see Fig.~\ref{fig:illustration}), where $\mathbf{c}$ is defined in analogy to $\mathbf{g}$ as the vector containing the coefficients $c_j$. 
For example, for the compliant sequence of Fig.~\ref{fig:AnalogDynamics_experiment}(a) below, we obtain $\mathcal{D}=0.0068$. 
This protection energy scale has to be compared to the spread of the energy eigenfunctions, which is roughly given by the norm of $H_0 + \lambda H_1$ and which for a generic many-body system is extensive in system size. Thus, for sufficiently large systems, one will always have overlap of undesired gauge sectors with the target sector $\mathbf{g}=\mathbf{0}$. Nevertheless, for errors consisting of terms with bounded spatial support such as in Fig.~\ref{fig:AnalogDynamics_experiment}(a) an exponentially long gauge protection is still assured for large but constant $V$, thanks to the theorem discussed above.

We can intuitively understand the gauge protection by going into an interaction picture with respect to $VH_G$. The full time-evolution operator then becomes
\begin{align}
	U(t)=&\,e^{-iHt}=e^{-iVH_Gt}\tilde{U}(t),\\\label{eq:Utilde}
	\tilde{U}(t)=&\,\mathcal{T}\big\{e^{-iH_0t-i\lambda\int_0^t \d\tau \lambda H_1(\tau)}\big\},
\end{align}
with $\lambda H_1(t)=e^{iVH_Gt}\lambda H_1e^{-iVH_Gt}$. 
We can compare the form of this time-dependent Hamiltonian with the projector onto the $\mathbf{g}=\mathbf{0}$ subspace, which for the $\mathrm{U}(1)$ LGT under consideration here can be written as ${P}_\mathbf{0}\propto \prod_j \int \d\alpha_j e^{-i\alpha_j G_j}$, i.e., the projector integrates over all possible gauge transformations \cite{Lamm2020}. Since in Eq.~\eqref{eq:Utilde} above, the different $G_j$ all rotate at fast but different frequencies, averaging over the slow timescales given by the system dynamics generates an effective projection onto the target gauge sector, where $V c_j t$ effectively assumes the role of the transformation angle $\alpha_j$. Said differently, the fast frequency $V$ rotates $H_1$ away. By choosing the $c_j$ in an incommensurate manner, it is ensured that each generator rotates independently of the others.

As final remarks, if a protection is desired for another target gauge sector $\textbf{g}^\star$, the above condition simply needs to be adjusted to $\sum_j c_j(g_j-g_j^\star)=0$ $\mathrm{iff}$ $\textbf{g}=\textbf{g}^\star$. 
In Appendix~\ref{sec:QZE}, we moreover use the ``continuous'' quantum Zeno effect (QZE) \cite{facchi2002quantum,burgarth2019generalized} to demonstrate that the protection term can be simplified [i.e., need not fulfil Eq.~\eqref{eq:c_j}] if the aim is just to protect gauge invariance to limited experimentally accessible times that are polynomially rather than exponentially large in $V$. Finally, $\tilde{H}_G$ as well as the generators of $\mathrm{Z}_2$ gauge theories (such as the stabilizers used for energy-gap protection \cite{Jordan2006,Young2013}) equally well fulfil the two main ingredients of the theorem: their spectrum is integer and $\Pi_{n=0}$ projects only onto the states with $\mathbf{g} =\mathbf{0}$ (for $\mathrm{Z}_2$ gauge theories, one needs to include a constant, irrelevant shift). Thus, we can immediately extend the theorem to the two-body protection $V\tilde{H}_G= V\sum_j G_j^2$, see Eq.~\eqref{eq:FullProtection}, as well as the energy-gap protection discussed  in Sec.~\ref{sec:DD-EGP}. In this way, these previously proposed protection schemes benefit from the same protective power of the above theorem.

\section{Numerical Results}\label{sec:results}
To substantiate these analytical considerations, this section presents numerical benchmarks for the gauge-violation dynamics in potential analog quantum simulators with continuous time evolution as well as digital circuits. The toolkits used for these results are QuTiP \cite{Johansson2012,Johansson2013} (analog) and Cirq (digital). 

We illustrate our theorem using as model the $\mathrm{U}(1)$ QLM defined by Eqs.~\eqref{eq:H0} and~\eqref{eq:Gj}. We prepare our initial state $\ket{\psi_0}$ in the gauge-invariant sector $G_j\ket{\psi_0}=0$, $\forall j$ (i.e., the sector $\mathbf{g}=\mathbf{0}$), and subsequently quench it at $t=0$ with the Hamiltonian $H=H_0+\lambda H_1+V H_G$ as per Eq.~\eqref{equ:effectiveHamiltonian}. 
Thus, the time-evolved wave function $\ket{\psi(t)}=U(t)\ket{\psi_0} = \exp[-i(H_0+\lambda H_1+V H_G)t]\ket{\psi_0}$ will in general no longer reside only in the initial gauge-invariant sector. The resulting violation in Gauss's law can be quantified by
\begin{align}\label{eq:violation}
\varepsilon(t)=\frac{1}{L}\sum_{j=1}^L\bra{\psi(t)}G_j^2\ket{\psi(t)}.
\end{align}
In what follows, we will compare the gauge violation for sequences that comply with Eq.~\eqref{eq:c_j} and sequences that do not, and we will compare the results with a quench using $H=H_0+\lambda H_1+V \tilde{H}_G$, with the two-body protection given by Eq.~\eqref{eq:FullProtection}.

\subsection{Analog quantum simulator with continuous time evolution}
\begin{figure}
	\centering
	\includegraphics[width=.48\textwidth]{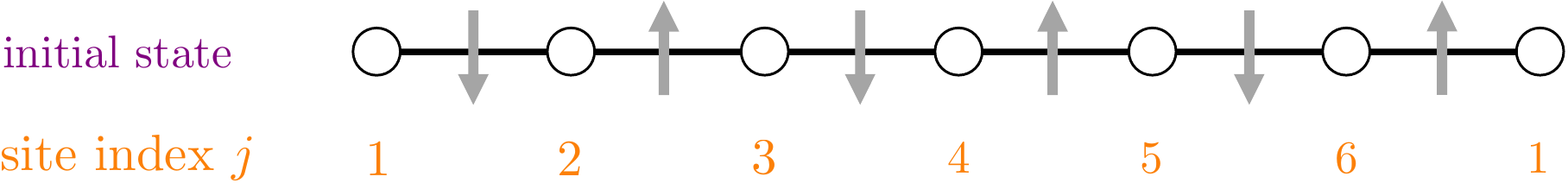}\\
	\includegraphics[width=.48\textwidth]{{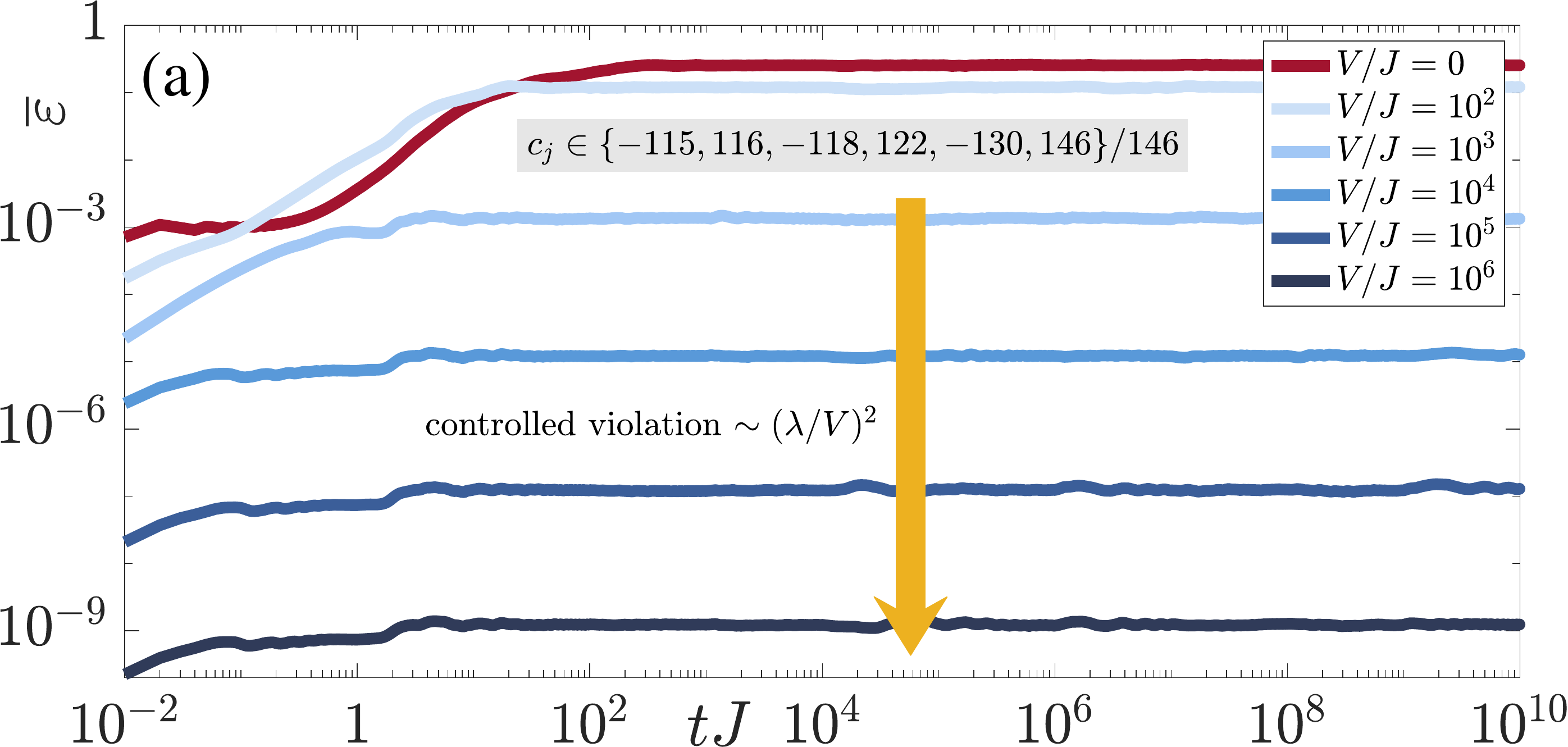}}\\
	\includegraphics[width=.48\textwidth]{{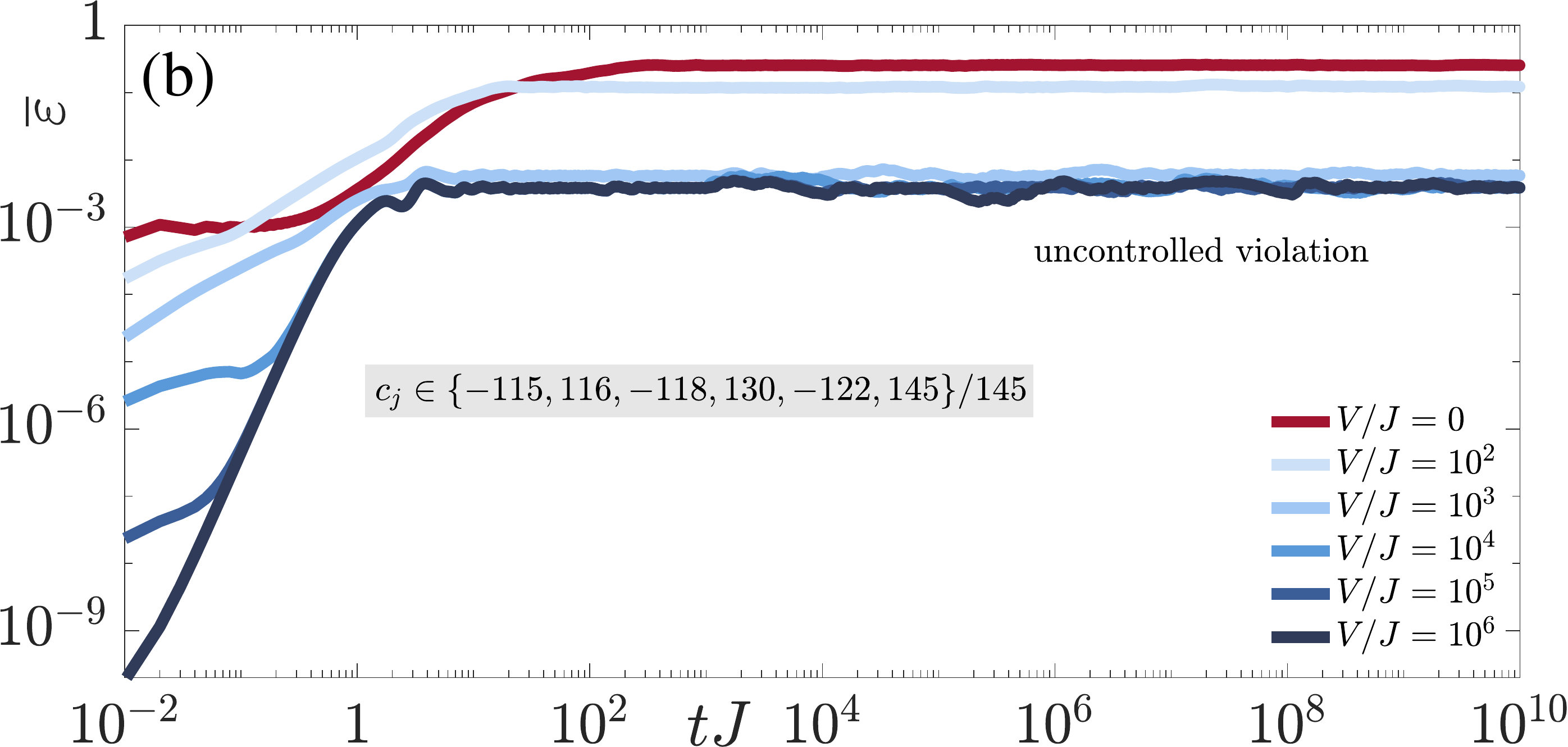}}\\
	\includegraphics[width=.48\textwidth]{{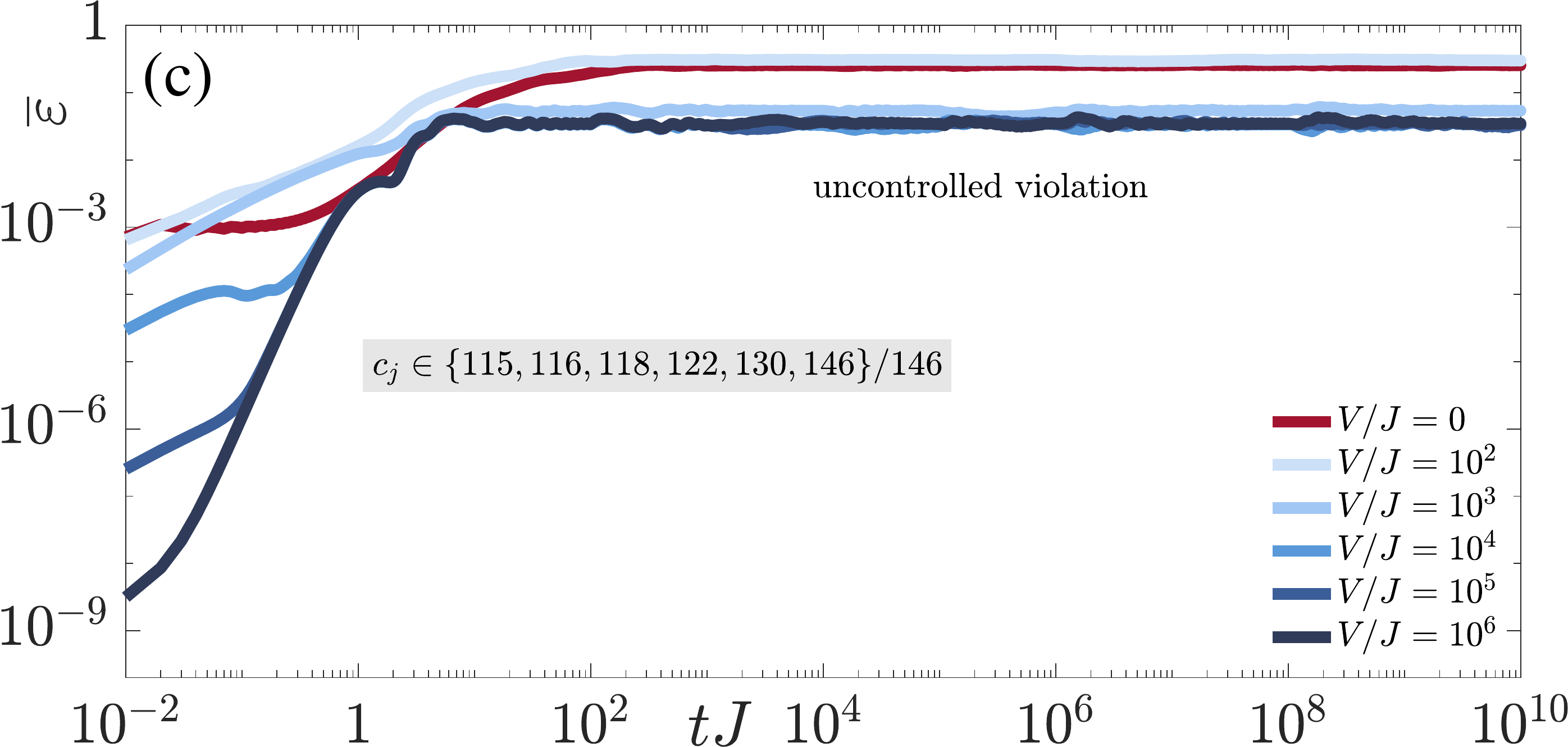}}
	\caption{(Color online). Spatiotemporally averaged gauge-invariance violation with the extreme nonlocal error $H_1$ of Eq.~\eqref{eq:H1} at gauge-breaking strength $\lambda=0.05J$, and under the single-body protection of Eq.~\eqref{eq:LinearProtection} for various values of the protection strengths $V$ (see legends). The initial state is drawn on top. (a) With a compliant sequence $\{c_j\}$, the gauge violation is suppressed $\sim(\lambda/V)^2$ for sufficiently large $V$, thereby bringing the dynamics perturbatively close to that of a renormalized version of the ideal gauge theory. (b) A noncompliant sequence provides only limited protection where the violation cannot be suppressed beyond a certain value, despite this sequence being very similar to the one in (a). (c) Similarly, the protection is also limited when the staggering in the compliant sequence is removed. Even though this might indicate a large sensitivity of the protection to details of the sequence, we show in Fig.~\ref{fig:AnalogDynamics_experiment} that some noncompliant sequences can still provide reliable protection for typical local gauge-breaking errors occurring in experimental setups.}
	\label{fig:AnalogDynamics}
\end{figure}

Typical errors in an analog quantum simulator experiment, such as with ultracold atoms in optical lattices, would violate the so-called assisted matter tunneling or gauge flipping of the first term in Eq.~\eqref{eq:H0} \cite{Mil2019}. However, some proposals also involve nonlocal gauge-breaking terms \cite{Yang2016}. Here, we consider as a worst-case scenario nonlocal error terms that guarantee the system is driven into all possible gauge-invariant sectors (we present results for only local errors further below). Specifically, we choose the gauge invariance-breaking term
\begin{align}\nonumber
H_1&=\sum_{j=1}^L \big(\tau_{j,j+1}^+ + \sigma_{j}^-\sigma_{j+1}^- + \mathrm{H.c.}\big)\\\label{eq:H1}
&+\sum_{\xi=\pm1}\prod_{j=1}^L\big(\mathds{1}_j+\xi\sigma^x_j\big)(\mathds{1}_{j,j+1}+\xi\tau^x_{j,j+1}\big).
\end{align}
We prepare our system in a staggered gauge-link configuration, with odd (even) links pointing down (up), and with all matter sites empty (see Fig.~\ref{fig:AnalogDynamics}, top), such that the resulting initial state $\ket{\psi_0}$ lies in the sector $\mathbf{g}=\mathbf{0}$. 

\begin{figure}[!ht]
	\centering
	\includegraphics[width=.48\textwidth]{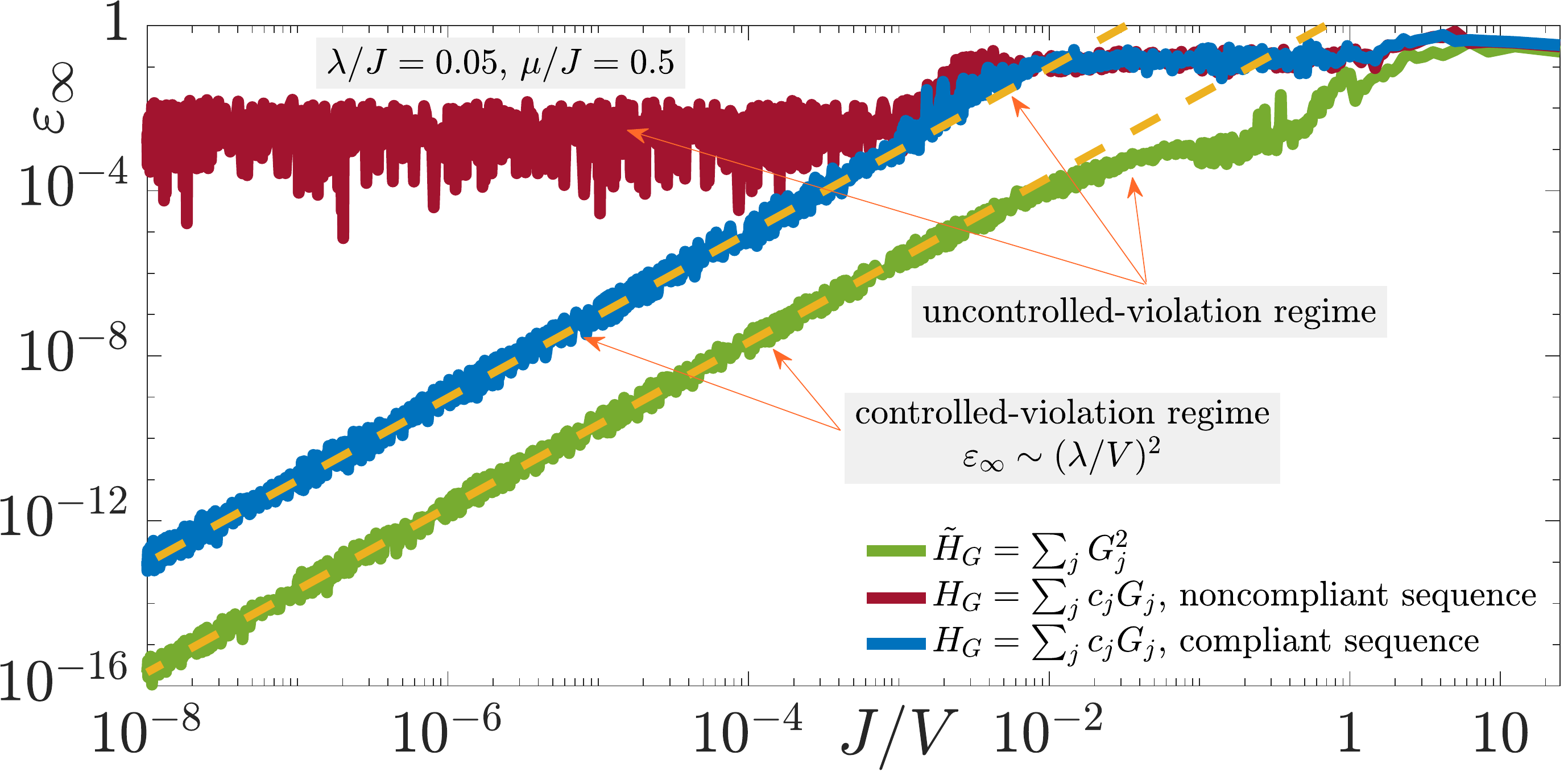}
	\caption{(Color online). Infinite-time gauge violation in gauge-theory dynamics at mass $\mu=0.5J$ with inherent gauge-breaking errors given by Eq.~\eqref{eq:H1} at breaking strength $\lambda=0.05J$, comparing the effect of the two-body energy penalty Eq.~\eqref{eq:FullProtection} (green curve), a single-body energy penalty with the compliant sequence $c_j\in\{-115,116,-118,122,-130,146\}/146$ (blue), and a single-body penalty with the noncompliant sequence $c_j\in\{-115,116,-118,130,-122,145\}/145$ (red). The two-body and compliant-sequence single-body penalties exhibit two distinct regimes: the first one is characterized by an uncontrolled violation when $V$ is too small, and the second regime exhibits a controlled gauge violation at large enough $V$ that scales $\sim(\lambda/V)^2$. In contrast, the noncompliant energy penalty displays only uncontrolled error behavior, which leads to a minimum violation that does not improve upon further increasing $V$. See Appendix~\ref{sec:AnalogSM} for similar results at different values of $\lambda$ and $\mu$ (Fig.~\ref{fig:AnalogScanSM}) and when starting in a different initial state (Fig.~\ref{fig:OtherInitial}).}
	\label{fig:AnalogScan}
\end{figure}

\begin{figure*}[!ht]
	\centering
	\includegraphics[width=.48\textwidth]{{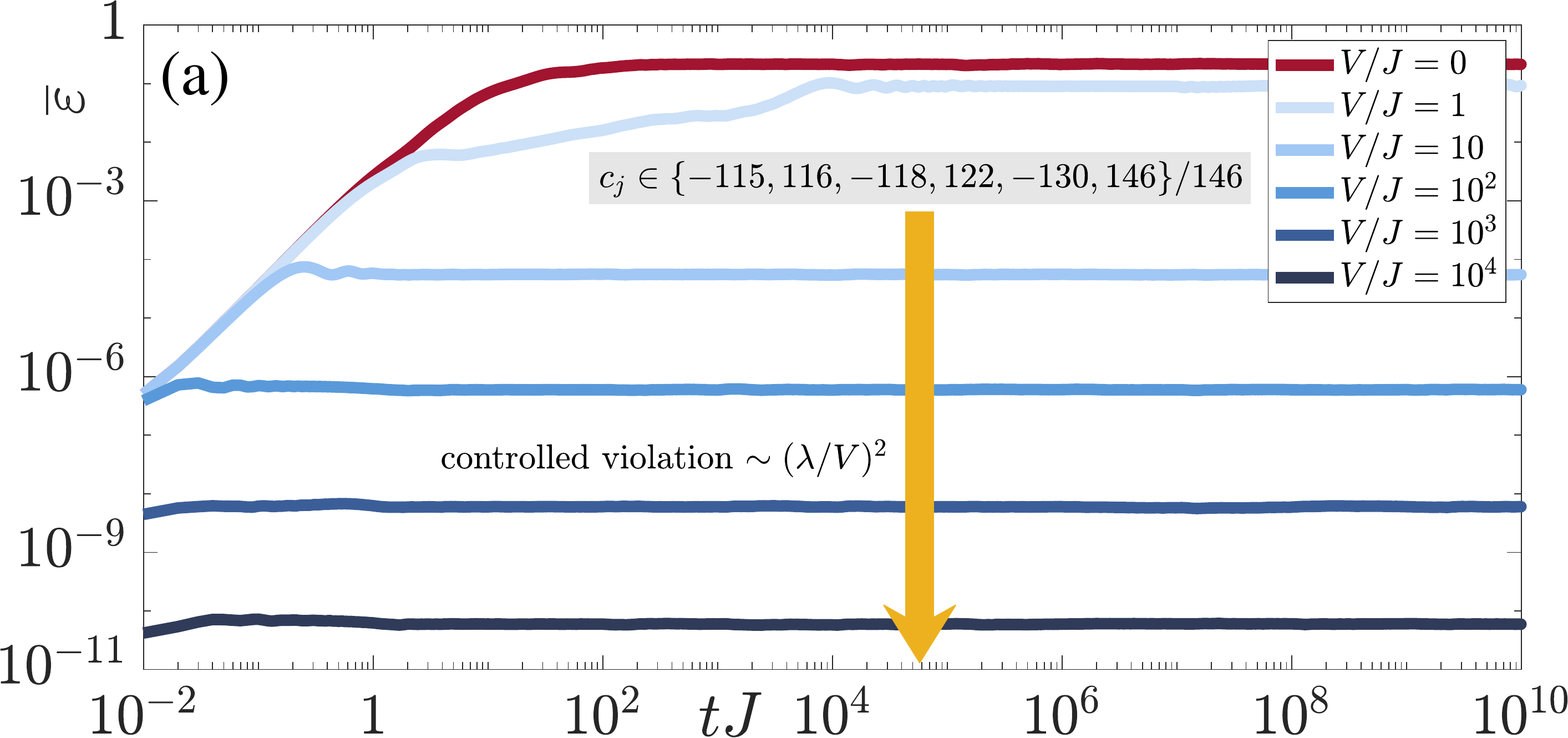}}\quad
	\includegraphics[width=.48\textwidth]{{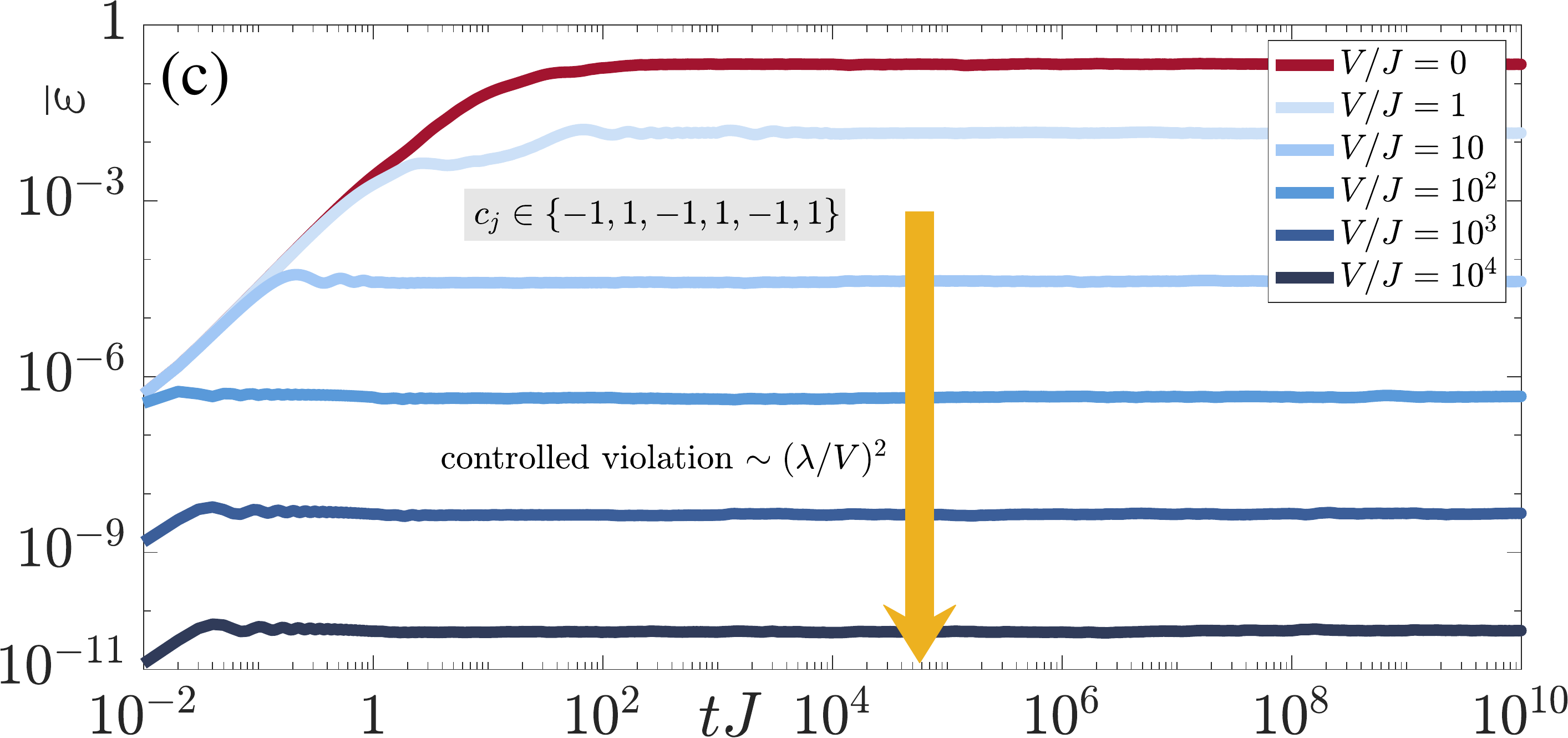}}\quad\\
	\includegraphics[width=.48\textwidth]{{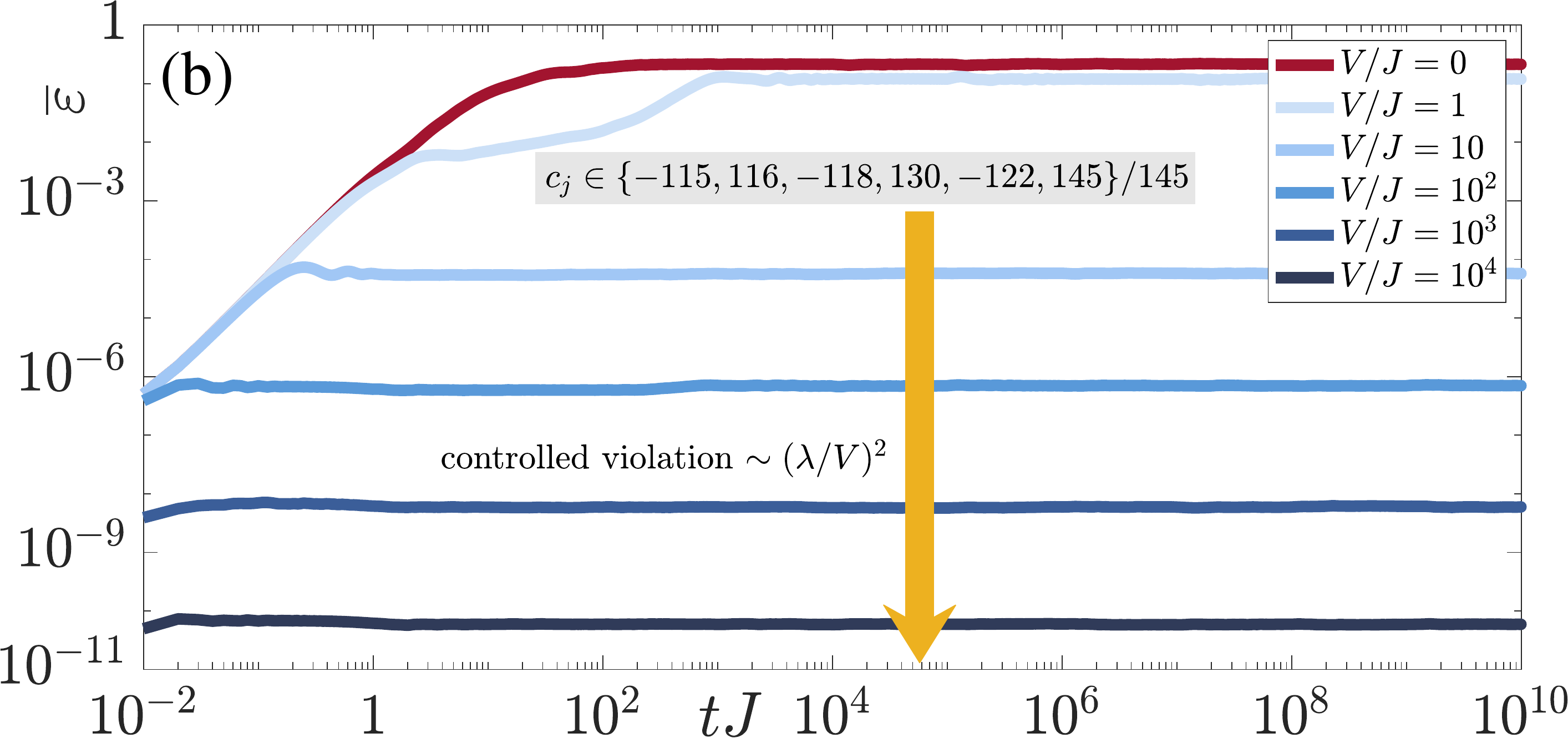}}\quad
	\includegraphics[width=.48\textwidth]{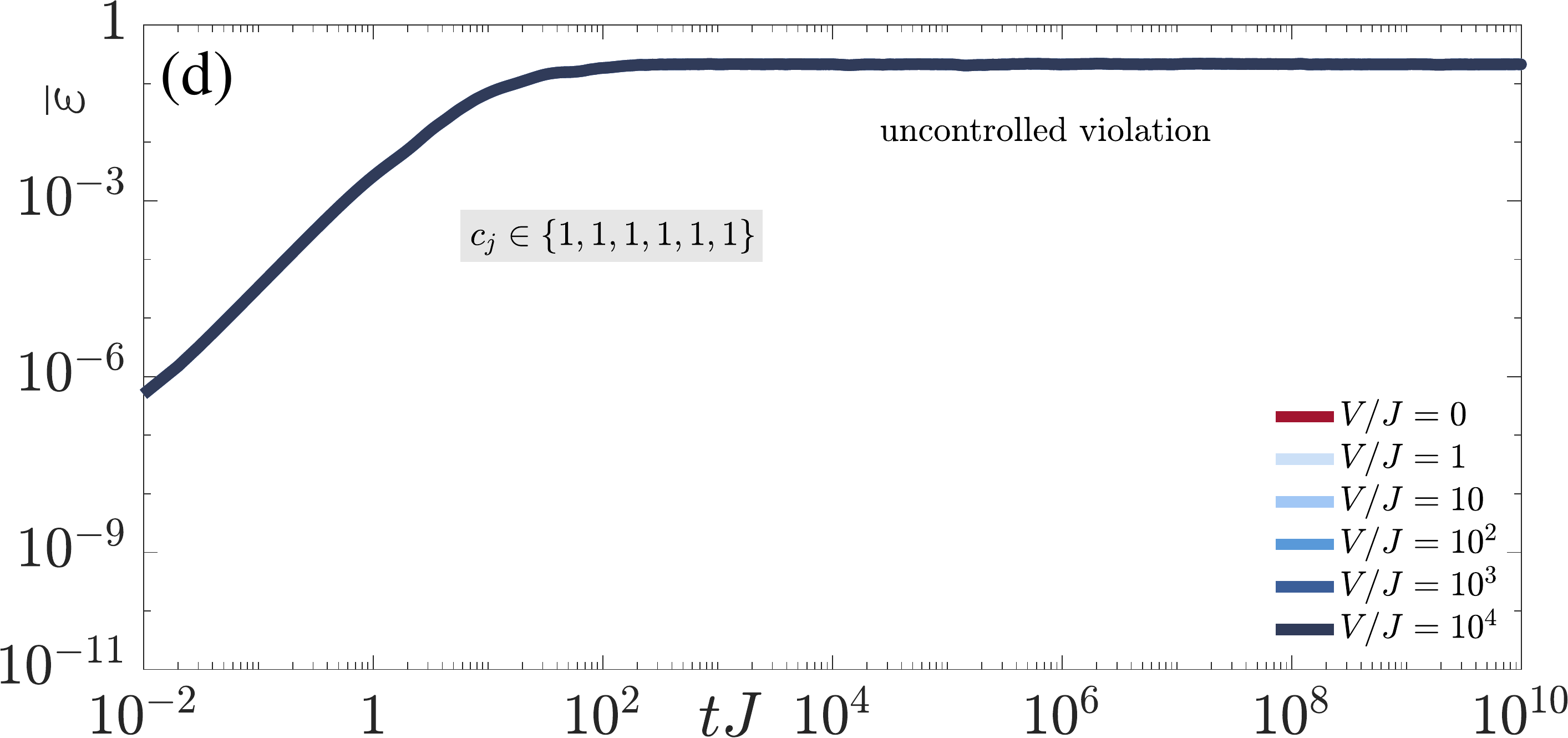}\quad
	\caption{(Color online). Similar to Fig.~\ref{fig:AnalogDynamics}, but with the local gauge-breaking term $H_1=\sum_j(\tau^x_{j,j+1}+\sigma^+_j\sigma^+_{j+1}+\sigma^-_j\sigma^-_{j+1})$. (a) As expected, the compliant sequence leads to excellent retention of gauge invariance. (b) Surprisingly, also the noncompliant sequence, which fails to protect against the gauge-breaking term of Eq.~\eqref{eq:H1}, provides very reliable protection. (c) Even more, in the present case already a simple staggered sequence of equal-magnitude coefficients can offer reliable gauge protection against gauge violations. (d) A simplistic sequence of equal terms will nevertheless still fail.}
	\label{fig:AnalogDynamics_experiment}
\end{figure*}

The running temporal average $\overline\varepsilon(t)=\int_0^t\d s\,\varepsilon(s)/t$ of the gauge violation in Eq.~\eqref{eq:violation} is shown in Fig.~\ref{fig:AnalogDynamics}, for gauge-breaking strength $\lambda=0.05J$ and various values of the protection strength $V$; see Appendix~\ref{sec:specs} for results on the temporally nonaveraged violation of Eq.~\eqref{eq:violation}. The compliant sequence employed in Fig.~\ref{fig:AnalogDynamics}(a) ensures a controlled suppression $\sim(\lambda/V)^2$ of the gauge violation at sufficiently large $V$, bringing the dynamics to that of the ideal gauge theory in the limit $V\to\infty$. In other words, the compliant sequence at sufficiently large $V$ allows one to extract from degenerate perturbation theory dynamics perturbatively close to that of a renormalized version of the ideal gauge theory. Therefore, the behavior is qualitatively identical to the case with two-body protection, i.e., when the energy penalty terms are quadratic in Gauss's-law operators \cite{Halimeh2020a}, but with the crucial advantage that now the protection term is linear in the $G_j$, i.e., requires only single-body terms. Importantly, the violation remains controlled over all simulated times, which go to values even beyond the shown maximal time of $t=10^{10}/J$. 

The picture drastically changes when the sequence does not comply with Eq.~\eqref{eq:c_j}, as shown in Fig.~\ref{fig:AnalogDynamics}(b,c), where no matter how large $V$ is, the gauge violation will not improve beyond a certain finite minimum value. The dynamics is thus no longer perturbatively close to the ideal gauge-invariant theory renormalized. Despite the similarity of the compliant and noncompliant sequences in Fig.~\ref{fig:AnalogDynamics}, they generate a strongly different associated dynamics. One may naively expect that this means the compliant sequence requires high accuracy in its implementation in order to \textit{exactly} satisfy Eq.~\eqref{eq:c_j}. However, as we show below, for typical experimental errors that are local, the compliant sequence is merely a sufficient but not a necessary condition to achieve controlled gauge violation. 

In Fig.~\ref{fig:AnalogScan}, we plot the infinite-time gauge violation as a function of $J/V$ (for fixed gauge-breaking strength $\lambda=0.05J$), comparing the different protection schemes. In congruence with the results of Fig.~\ref{fig:AnalogDynamics}, the compliant-sequence single-body protection offers the same two-regime picture as its two-body counterpart, albeit the gauge violation is unsurprisingly smaller with the two-body protection. Nevertheless, in both cases at sufficiently large $V$, the violation is suppressed $\sim(\lambda/V)^2$, allowing a perturbative reconstruction of the ideal gauge-theory dynamics through a controlled extrapolation towards $\lambda/V\to 0$. In the case of a noncompliant sequence, the gauge violation is shown to be suppressed only down to a finite minimum value regardless of how large $V$ is. ``Infinite time'' in Fig.~\ref{fig:AnalogScan} refers to $t=10^{10}/J$, but we have checked that our conclusions remain qualitatively the same for much larger evolution times.

The results of Figs.~\ref{fig:AnalogDynamics} and~\ref{fig:AnalogScan} may lead to the false impression that the compliant sequence must be engineered with great accuracy in order to achieve controlled violation. However, this is only true in the case of extreme gauge-breaking terms where $H_1$ takes a nonlocal form such as in Eq.~\eqref{eq:H1}. In realistic settings, dominant gauge-breaking terms are usually those stemming from unassisted matter coupling or gauge flipping \cite{Poppitz2008,Mil2019,Yang2020}. Here, we model these by
\begin{align}\label{eq:expH1}
H_1=\sum_j\big(\tau^x_{j,j+1}+\sigma^+_j\sigma^+_{j+1}+\sigma^-_j\sigma^-_{j+1}\big).
\end{align}

Results for the associated time evolution of the gauge violation are shown in Fig.~\ref{fig:AnalogDynamics_experiment}(a,b) for the same compliant and noncompliant sequences used in Fig.~\ref{fig:AnalogDynamics}(a,b). The compliant sequence again performs remarkably well, but, intriguingly, in the present case the sequence not compliant with Eq.~\eqref{eq:c_j} also works reliably, despite offering no control in the case of the extreme error of Eq.~\eqref{eq:expH1}. Even more, as shown in Fig.~\ref{fig:AnalogDynamics_experiment}(c), another noncompliant sequence that involves coefficients of equal magnitude but alternating sign also shows excellent control of the gauge violation. Nevertheless, equal coefficients of the same sign do not allow one to control the gauge violation, as shown in Fig.~\ref{fig:AnalogDynamics_experiment}(d). The Appendix~\ref{sec:AnalogSM_scan_expError} contains scans similar to Fig.~\ref{fig:AnalogScan} of the infinite-time gauge violation as a function of $J/V$. 

To illustrate the power of the linear gauge protection theorem, we give an estimate for the protection strength based on Eqs.~\eqref{eq:Vge} and~\eqref{eq:nstar} for the local error term Eq.~\eqref{eq:expH1} and the extreme error term Eq.~\eqref{eq:H1}. For convenience, we still assume $\lambda=0.05J$ (though the gauge protection does not depend on $\lambda$ being perturbatively small compared to $J$). In such a parameter setting, Eq.~\eqref{eq:nstar} has more restrictions on the minimal protection strength $V_{\min}$. By finding the $\kappa_0$ that minimizes $V_0$, we find the value of $V_{\min}$ above which exponentially long gauge protection is guaranteed. First, considering the local error, the energy scale $V_0$ is dominated by $H_0$ in the chosen parameter setting, which gives $V_0 \approx 3000J$ and $V_{\min}\approx 2000J$. As for the extreme error, strictly speaking the interaction range is infinite, hence it does not belong to the applicable range of the theorem of Ref.~\cite{Abanin2017} in the thermodynamic limit. However, this is not an issue for lattices of finite size. For the system we consider in this paper, $L=6$ matter sites, the energy scale for the extreme error is $V_0 \approx 8000J $, which leads to $V_{\min} \approx 5000J$. The above estimation is for unnormalized $c_j$ and $V$ [see discussion below Eq.~\eqref{eq:c_j}]. As a comparison, Figs.~\ref{fig:AnalogDynamics} and~\ref{fig:AnalogDynamics_experiment} show that the unnormalized protection strength starts to work at $V_{\min}\approx 0.1J$ and $V_{\min}\approx J$ for both the local and extreme (nonlocal) errors, respectively, at least for the finite system sizes considered here.  As these estimates for the protection strength show, the requirements on parameters in an actual quantum simulation are much more feasible than what the theory predicts.
	
\subsection{Digital circuit with discrete time evolution}
\label{sec:digitalCircuit}

\begin{figure}[t]
	\centering
	\includegraphics[width=\columnwidth]{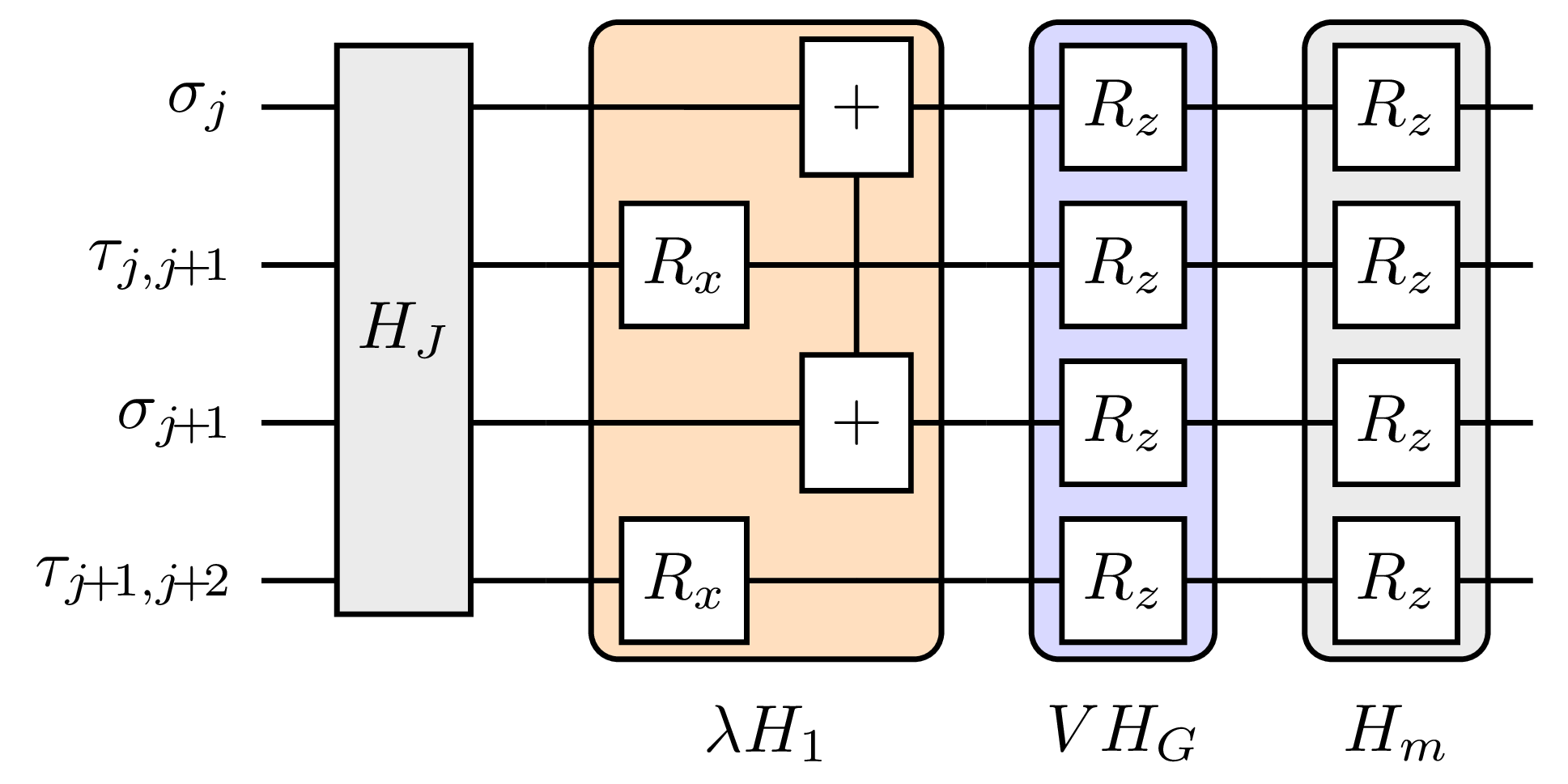}
	\caption{Elementary unit for one Trotter step of the quantum circuit. $\sigma_j$ ($\tau_{j,j+1}$) denotes a qubit representing a matter (gauge) field at matter site $j$ [gauge link ($j,j+1$)]. Single-qubit rotations around qubit axis $\alpha$ are labelled by $R_{\alpha}=\exp(-i\sigma^\alpha\phi/2)$ for qubits representing matter sites and $R_{\alpha}=\exp(-i\tau^\alpha\phi/2)$ for qubits representing gauge links, where the angle $\phi$ is defined by the relative weight of the corresponding term in the Hamiltonian. ``$+$'' denotes the unitary gate $\exp[-i(\sigma^+_j\sigma^+_{j+1}+\text{H.c.})\lambda\delta t]$ with Trotter time step $\delta t$.  The implementation of $VH_G$ and $H_m$ can be combined in one layer of single-qubit $z$-rotations.}
	\label{fig:QuantumCircuit}
\end{figure}

\begin{figure}[!ht]
	\centering
	\includegraphics[width=.48\textwidth]{{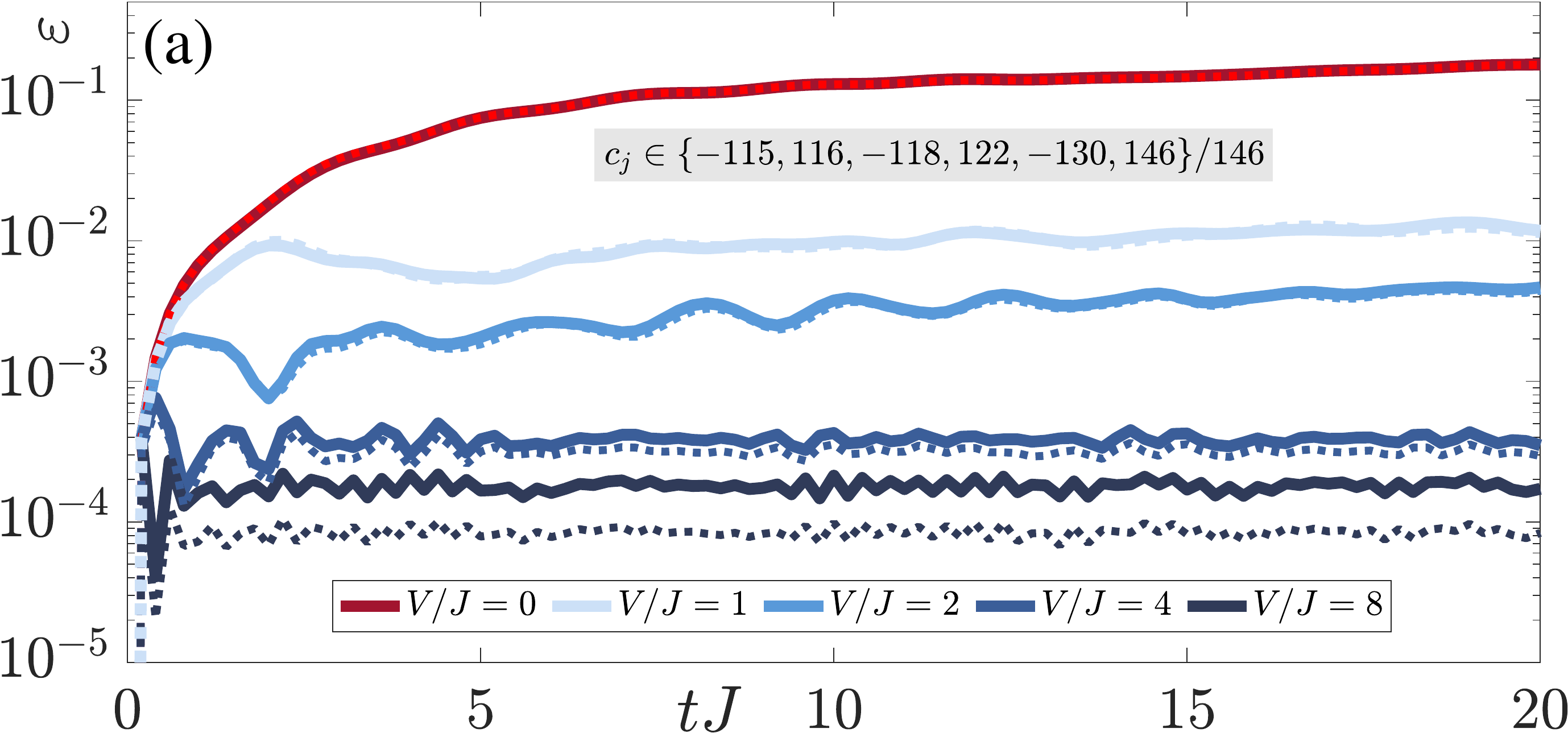}}\\
	\includegraphics[width=.48\textwidth]{{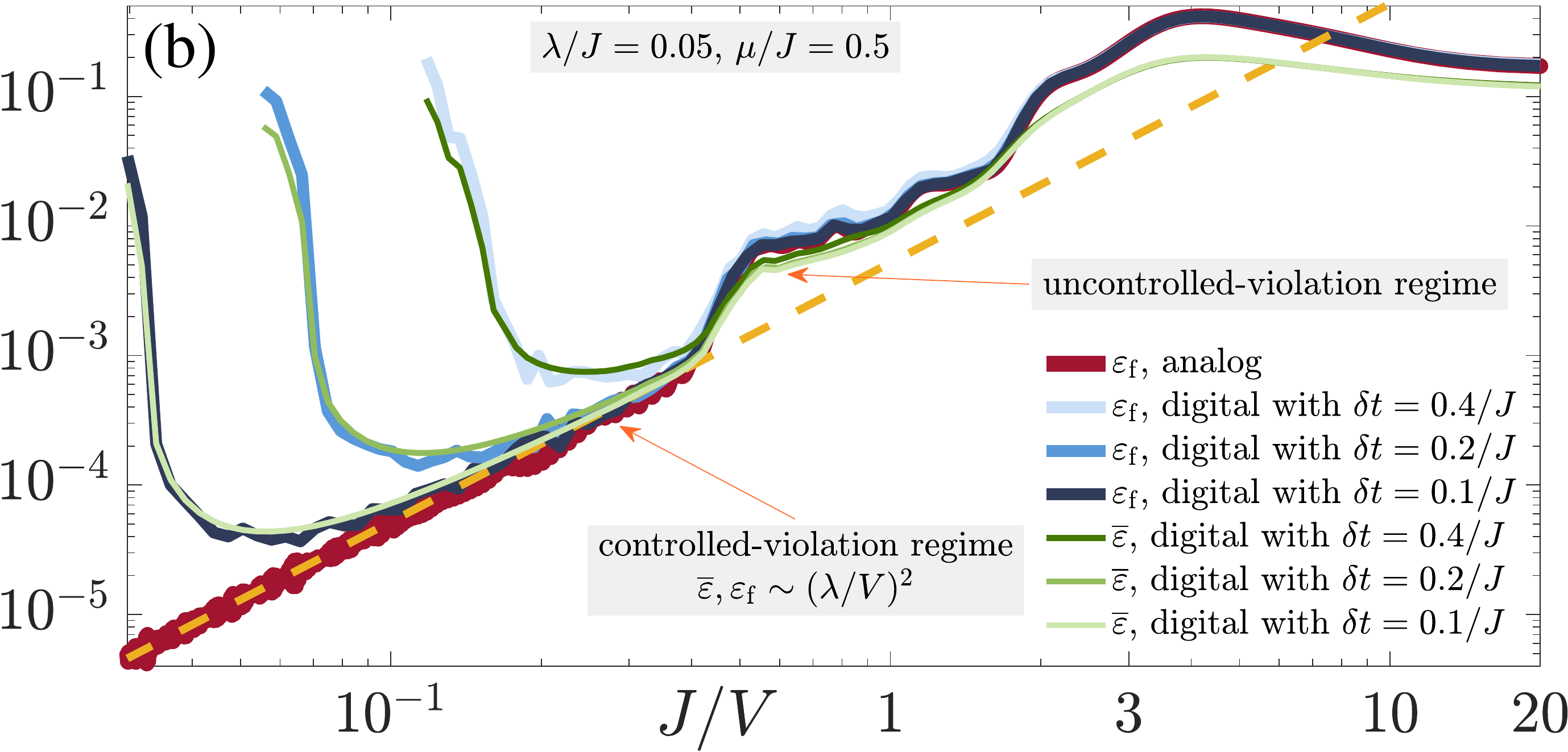}}
	\caption{(Color online). (a) Numerical benchmarks of gauge violation dynamics in a digital circuit, for a Trotter time step $\delta t=0.2/J$ at various values of gauge-protection strength $V$, with $\mu/J=0.5$, $\lambda/J=0.05$, and $L=6$ matter sites. The corresponding numerical data for the analog quantum simulator are shown in dotted lines of the same color. (b) The gauge violation at final time $t_\mathrm{f}=20/J$ in the analog quantum simulator and in the digital circuit for various Trotter time steps (see legend). The gauge violation reaches a broad minimum around the ideal protection strength of $V_\mathrm{ideal}\approx\pi/(2\overline{c}\delta t)-\xi$, see Eq.~\eqref{eq:Videal}.}
	\label{fig:Digital}
\end{figure}

In this section, we numerically benchmark the gauge protection under $H = H_0+\lambda H_1+VH_G$ of Eq.~\eqref{equ:effectiveHamiltonian} in a digital circuit. For this purpose, we assume a linear arrangement of qubits that alternately represent matter and gauge fields, and we choose the same initial state as in the previous section (see top of Fig.~\ref{fig:AnalogDynamics}). 
In the simulated digital quantum circuit, sketched in Fig.~\ref{fig:QuantumCircuit}, the time evolution generated by the different parts of the Hamiltonian $H$ is implemented in the separate layers $\exp(-iH_J\delta t)$, $\exp(-i\lambda H_1\delta t)$, $\exp(-iVH_G\delta t)$, and $\exp(-iH_m\delta t)$ with $\delta t$ the Trotter time step. Here, it is beneficial to split the $\mathrm{U}(1)$ Hamiltonian in Quantum Link Formalism $H_0$ of Eq.~\eqref{eq:H0} into the kinetic energy term coupling matter and gauge fields $H_J=J\sum_j\left(\sigma_j^-\tau_{j,j+1}^+\sigma_{j+1}^-+\mathrm{H.c.}\right)$ and the fermionic rest mass term $H_m=\frac{\mu}{2}\sum_j\sigma_j^z$. With this, both $H_m$ and $VH_G$ can be implemented by single-qubit $z$-rotations $R_{z,j}(\phi)=\exp(-i\sigma_j^z\phi/2)$ for qubits representing matter sites and $R_{z,(j,j+1)}(\phi)=\exp(-i\tau^z_{j,j+1}\phi/2)$ for qubits representing gauge links, where the angle $\phi$ is defined by the relative weight in the Hamiltonian, e.g., $\exp(-iH_m\delta t)=\bigotimes_j \exp(-i \sigma_j^z\mu\delta t/2)=\bigotimes_jR_{z,j}(\mu\delta t)$. As $H_m$ and $VH_G$ commute, the single-body gauge-protection Hamiltonian $VH_G$ can be implemented in combination with the single-qubit rotations of the $H_m$ layer without increasing gate depth. 
Since we are specifically interested in a controlled study of gauge violation, we assume an exact implementation of $H_J$ and add a gauge-breaking term $\lambda H_1$ by hand, which mimics imperfectly calibrated gates and other systematic gauge-violations that may occur in a realistic implementation. We choose the local gauge-breaking Hamiltonian $\lambda H_1$ of Eq.~\eqref{eq:expH1} that is split into single-qubit $x$-rotations $R_{x,(j,j+1)}(2\lambda\delta t)=\exp(-i\tau^x_{j,j+1}\lambda\delta t)$ realising the gauge-flipping term and the two-qubit gate $\exp[-i(\sigma^+_j\sigma^+_{j+1}+\text{H.c.})\lambda\delta t]$ realising the matter coupling. 
In what follows, we choose a sequence of $c_j$ that complies with Eq.~\eqref{eq:c_j}, though---as in the continuous-time calculations of Figs.~\ref{fig:AnalogDynamics_experiment} and~\ref{fig:AnalogScan_experiment}---a simple sequence of constant magnitude and alternating sign already yields controlled protection for the used local error term.

As Fig.~\ref{fig:Digital}(a) shows, the gauge violation $\varepsilon$ can be efficiently suppressed by choosing $V>0$. For $V/\lambda$ large enough, we observe the controlled-error regime where the gauge violation is suppressed $\sim(\lambda/V)^2$, see Fig.~\ref{fig:Digital}(b). Moreover, there is a scale of optimal gauge protection, $V_{\mathrm{ideal}}$. For $V$ sufficiently smaller than $V_{\mathrm{ideal}}$, the digital error suppression coincides with the continuous-time simulations of the preceding section (up to Trotter errors), whereas above it the digital suppression of gauge violation begins to deteriorate. 
We find this ideal gauge-protection strength to be given by
\begin{equation}
V_{\mathrm{ideal}}\approx\frac{\pi}{2\overline{c}}\delta t^{-1}-\xi,
\label{eq:Videal}           
\end{equation}
where $\overline{c}$ is the spatial average of the absolute values of the coefficients $c_j$, which due to the normalization discussed following Eq.~\eqref{eq:c_j} is smaller than but on the order of unity.
Intuitively, the first term is the protection strength above which the $z$-rotation angle of the qubits on the Bloch sphere exceeds the order of $\pi$, i.e., the protection per Trotter step starts to actually become weaker. 
This value acquires a small correction $\xi$ that depends on $\mu$ and the microscopic details of the gauge-breaking term $\lambda H_1$. 
As shown in Appendix~\ref{sec:VidealSM}, when rescaling based on Eq.~\eqref{eq:Videal} the results for various $\delta t$ collapse onto each other. 
Moreover, as seen in Fig.~\ref{fig:Digital}(b), the achievable $\varepsilon$ attains a broad minimum over $V$ centered around $V_{\mathrm{ideal}}$, meaning no experimental fine tuning is needed to reach the optimal gauge protection. 

It is remarkable that the gauge protection works so reliably also in the digital implementation, as the theorem in Sec.~\ref{sec:theorem} is derived for continuous time evolution. Nevertheless, as we discuss in detail in Appendix~\ref{sec:QZE}, in the case of Trotterized time evolution the QZE for coherent dynamics ensures $VH_G$ protects gauge invariance against unitary errors at least up to polynomially long times. Even more, as we have seen in the numerics, already for moderately large $V$ we find an approximately constant level of gauge suppression up to the simulated times of $20/J$.

\section{Discussion} \label{sec:discussion}
In this section, we put our protection framework in context of previous results. In particular, we formally relate the proposed method with the frameworks of dynamical decoupling and energy-gap protection by introducing an auxiliary, fictitious Higgs field. Moreover, we discuss how a recent cold-atom experiment can be reinterpreted as implementing a simplified noncompliant gauge-protection sequence.

	\subsection{Relation to dynamical decoupling and energy gap protection}\label{sec:DD-EGP}
	
	It is instructive to put our protection scheme into relation to the known techniques of energy gap protection (EGP), which uses time-independent suppression terms, and dynamical decoupling (DD), which relies on time-dependent sequences. 
	Both techniques have been proposed to provide error mitigation, e.g., by encoding logical qubits into stabilizer codes in the context of adiabatic quantum computing \cite{lidar_towards_2008,Young2013,vinci_nested_2016}. Such error-detecting or error-correcting codes can be understood as $\mathrm{Z}_2$ gauge theories where the stabilizers assume the role of the generators of a $\mathrm{Z}_2$ Gauss's law. 
	
	In the frameworks of DD and EGP, one is concerned with suppressing errors that occur through the interaction of a target system, typically the qubit register of a quantum computer, with the environment. The corresponding Hilbert spaces are $\mathcal{H_\mathrm{sys}}$ and $\mathcal{H_\mathrm{E}}$, respectively, and the Hilbert space of the composite system is given by the direct product $\mathcal{H}=\mathcal{H_\mathrm{sys}}\otimes \mathcal{H_\mathrm{E}}$. The full dynamics is governed by Hamiltonian $H=H_\mathrm{sys}\otimes \mathds{1}_\mathrm{E}+\mathds{1}_\mathrm{sys}\otimes H_\mathrm{E}+\lambda \sum_q O_\mathrm{sys}^{(q)}\otimes O_\mathrm{E}^{(q)}$. Here, $H_\mathrm{sys}$ and $H_\mathrm{E}$ govern the dynamics of target system respectively environment alone, which become coupled by the operators $O_\mathrm{sys}^{(q)}$ and $O_\mathrm{E}^{(q)}$ with overall strength $\lambda$. 
	
	The aim of DD and EGP is to suppress this coupling. To this end, one adds a control pulse $H_\mathrm{c}(t)\otimes \mathds{1}_\mathrm{E}$, assuming that (only) the target system can be actively manipulated, such that the full time evolution operator becomes $U_\mathrm{DD}(t)=\mathcal{T}e^{-i\int_0^t d \tau (H+H_\mathrm{c}(\tau)\otimes \mathds{1}_\mathrm{E})}$. 
	In the context of error-correcting codes, $H_\mathrm{c}$ typically consist of stabilizer operators \cite{Young2013}.  
	In the framework of EGP, one takes $H_\mathrm{c}$ to be constant, while DD employs suitably chosen, time-dependent control pulses $H_\mathrm{c}(t)\otimes \mathds{1}_\mathrm{E}$, typically assuming the control pulse to be cyclic with period $T_\mathrm{c}$, i.e., $U_\mathrm{c}(t)\equiv e^{-i\int_0^t d\tau H_\mathrm{c}(\tau)\otimes \mathds{1}_\mathrm{E}} = U_\mathrm{c}(t+T_\mathrm{c})$ (since the stabilizers all commute with each other, we can omit the time ordering prescription here). 
	In a rotating frame generated by $U_\mathrm{c}(t)$, one obtains $U_\mathrm{DD}(t)=U_\mathrm{c}(t)\tilde{U}_\mathrm{DD}(t)$, with 
	\begin{align}
	\tilde{U}_\mathrm{DD}(t)=&\,\mathcal{T}\big\{e^{-i \int_0^t d\tau \tilde{H}(\tau) }\big\},
	\end{align}
	and $\tilde{H}(t)=U_\mathrm{c}^\dagger(t)H U_\mathrm{c}(t)$. 
	
	To estimate the resulting dynamics, one may perform a Magnus expansion \cite{Blanes2009} of $\tilde{U}_\mathrm{DD}(N T_\mathrm{c})=e^{-i\sum_\ell \bar{H}^{(\ell)} N T_\mathrm{c}}$ \cite{Viola1999}, yielding a series of effective Hamiltonians $\bar{H}^{(\ell)}$ that describe the stroboscopic dynamics at each cycle. For example, the leading order is simply the time average $\bar{H}^{(1)}=\frac{1}{T_\mathrm{c}}\int_0^{T_\mathrm{c}}dt \tilde{H}(t)$. 
	Essentially the same reasoning can be applied to EGP when taking $H_\mathrm{c}(t)$ to be time-independent. The temporal periodicity of the control pulse is then simply given by a sine-wave function generated by $e^{-iH_\mathrm{c}t}$. 
	In such an effective description, $\bar{H}_\mathrm{sys}$ may get renormalized. More importantly in the present context, the coupling is modified to $\lambda \sum_q \bar{O}_\mathrm{sys}^{(q)}\otimes O_\mathrm{E}^{(q)}$. For a suitably chosen $H_\mathrm{c}$, $\bar{O}_\mathrm{sys}^{(q)}$ is averaged to zero, so to leading order the coupling between target system and environment is cancelled. 
	
	It may be tempting to try and reformulate our protection scheme in this framework, with target space $\mathcal{H_{\mathbf{g}=\mathbf{0}}}$ and undesired space $\mathcal{H_{\mathbf{g}\neq\mathbf{0}}}$. However, the full Hilbert space of the gauge theory takes the form of a direct sum $\mathcal{H}_\mathrm{GT}=\mathcal{H_{\mathbf{g}=\mathbf{0}}}\oplus \mathcal{H_{\mathbf{g}\neq\mathbf{0}}}$ rather than the direct product $\mathcal{H}=\mathcal{H_\mathrm{sys}}\otimes \mathcal{H_\mathrm{E}}$. We can nevertheless put the gauge theory in this framework by introducing auxiliary bosonic Higgs fields $\phi$ that assume the role of the environment, $\mathcal{H}=\mathcal{H}_\mathrm{GT}\otimes \mathcal{H}_\mathrm{Higgs}$. The gauge-breaking terms $\lambda H_1$ are then formally reinterpreted as the coupling between matter or gauge fields to the (fictitious) Higgs field. For example, the term $\lambda\sum_{j=1}^L \big(\tau_{j,j+1}^+ +\mathrm{h.c.}\big)$ appearing in Eqs.~\eqref{eq:H1} and~\eqref{eq:expH1} is then rewritten as $\lambda\sum_{j=1}^L \big( \tau_{j,j+1}^+ \otimes \phi_j^\dagger\phi_{j+1} +\mathrm{h.c.}\big)$ \cite{Poppitz2008,Bazavov2015}. 
	Since $H_G$ has an integer spectrum by construction, it fulfils the cyclic property, and thus $VH_G\otimes \mathds{1}_\mathrm{Higgs}$ assumes the role of the periodic control pulse $H_\mathrm{c}\otimes \mathds{1}_\mathrm{E}$. 
	Note that this is just a formal reinterpretation, the Higgs field is not actually being quantum simulated or added as an additional degree of freedom. Moreover, the Higgs field does not represent a dissipative, Markovian bath but rather generates a coupling between different gauge sectors. 
		
	Using this formulation, we can reinterpret our protection framework as a dynamical decoupling of the gauge theory from an auxiliary Higgs field---but with some important differences in terms of experimental requirements. 
	For example, previous proposals in the context of stabilizer gauge theories suffer from the necessity to add high-weight many-body terms \cite{Young2013}, while our framework requires only inexpensive single-body operators. 
	Moreover, DD proposals have discussed two ways to discard couplings that appear in higher orders of the Magnus expansion and which might deteriorate the target dynamics at polynomial time scales. First, when increasing the strength of the decoupling pulse with simulated time and system size, the Magnus expansion can always be shown to converge, enabling a controlled truncation of the series \cite{Blanes2009}. Second, $H_\mathrm{c}$ can be constructed through increasingly complex many-body terms that cancel $\bar{H}^{(\ell)}$ order by order \cite{Viola1999}. In our work, we show that such drastic requirement on the control Hamiltonian are unnecessary: decoupling of few-body error terms can be achieved for exponentially long times with a protection strength that remains constant in time and system size. Even more, for the $\mathrm{U}(1)$ gauge theory considered here, this can be achieved with simple single-qubit terms.

\subsection{Cold-atom implementations}\label{sec:coldatom}

The proposed protection scheme is directly relevant to ongoing cold-atom quantum simulations. For example, in a recent experiment \cite{Yang2020}, an optical superlattice has been designed in such a way as to impose energy penalties on the most salient gauge violations. The experiment distinguishes bosons on matter sites and on gauge links, described by bosonic operators ${b}_j$ and $b_{j,j+1}$, with associated number operators ${n}_j$ and $n_{j,j+1}$. Thanks to on-site interactions, an alternating chemical potential $\delta$ due to the superlattice, and a lattice tilt $\Delta$ due to gravity, the bosons are subject to the energy penalty $H_\mathrm{penalty}=\sum_j \{U [n_{j} (n_{j} +1) + n_{j,j+1} (n_{j,j+1} +1)]/2 + \delta n_{j,j+1} + \Delta [ j n_j + (j +1/2) n_{j,j+1}] \} $. With the generators of the target Gauss's law, $G_j= (-1)^j[(n_{j-1,j}+n_{j,j+1})/2 + n_{j} -1]$, the penalties can be rewritten as 
$H_\mathrm{penalty}= \sum_j \{ U [n_{j} (n_{j} +1) + n_{j,j+1} (n_{j,j+1} +1) - n_{j,j+1}/2 ]/2 - \mu n_j + c_j G_j \}$. 
For a large on-site interaction $U$, the first term $\propto U$ restricts the matter sites to occupations $0$ and $1$ and the gauge sites to occupations $0$ and $2$, enabling a mapping to the QLM given in Eq.~\eqref{eq:H0}. The second term $\propto \mu$ is mapped to the rest mass. Finally, within our framework, the third term is reinterpreted as a gauge protection consisting of a linear and a staggered term, $c_j = (-1)^j[\Delta j + (U-\delta + \Delta/2)]$. 

Although the coefficients do not satisfy the full condition of a compliant sequence as per the theorem of Sec.~\ref{sec:theorem}, the dynamics can still be protected by the QZE discussed in Appendix~\ref{sec:QZE}. 
The closest gauge sector $\textbf{g}$ degenerate with $\textbf{0}$ (i.e., the sector with the minimal $||\textbf{g}||^2>0$ such that $\sum_j c_j g_j=0$) takes the form $\textbf{g} =  (0,\ldots, 1, -1,  1, -1, 0,\ldots,0)$. Only a gauge-breaking term $H_1$ that acts on at least three matter/gauge field sites has the possibility to access this sector, for instance, $H_1 = \sum_j(\sigma_{j-1}^-\tau_{j,j+1}^+\sigma_{j+2}^- + \text{H.c.})$. Hence, even in this case the gauge invariance is protected up to a linear timescale when the perturbation term is sufficiently local. 
In a cold-atom experiment, $\Delta$ can be realized by gravity, a magnetic gradient, or a light shift. For an experiment of the type of Ref.~\cite{Yang2020} the maximal protection strength is restricted to $\Delta \approx 10 \mathrm{kHz}$ before errors due to higher bands become significant \cite{misc1}, and could thus be orders of magnitude larger than the most salient gauge violation that was suppressed to a level of $\lesssim 70\mathrm{Hz}$. A similar protection term could also be engineered in other cold-atom platforms, e.g., through AC-Stark shifts in the experiment of Ref.~\cite{Mil2019}. 

In the experiment of Ref.~\cite{Yang2020}, the penalty coefficients have been chosen \textit{ad hoc} to suppress the most salient errors of nearest-neighbor and next-nearest neighbor tunneling. As we see, they find an elegant reinterpretation in our framework, which thus also highlights a clear way forward to improve gauge protection in future experiments, e.g., by identifying the next subleading gauge-breaking terms along with sequences that protect against them.

\section{Conclusion}\label{sec:conclusion}
In summary, we have introduced the \textit{gauge protection theorem}: it proves reliable gauge invariance against coherent errors with bounded support up to exponentially long times and independent of system size, by using simple single-body terms proportional to the Gauss's-law generators. Each of these operators is weighted according to a \textit{compliant} sequence of coefficients such that their sum can be zero if and only if the quantum state resides in the target gauge sector, while other gauge sectors incur an energy penalty that serves as a \textit{single-body gauge protection}. As a consequence, the protection term generates an emergent global symmetry that within the target gauge sector acts in the same way as the local gauge symmetry.

Using numerical benchmark calculations, we have demonstrated the power of our method for near-future analog and digital quantum simulations of a $\mathrm{U}(1)$ gauge theory. Even in the presence of extreme nonlocal gauge-breaking terms, the single-body protection offers controlled gauge violation down to a perturbatively small level. Indeed, even though the theorem stipulates protection up to exponentially long times, we see that in our finite systems the gauge violation is suppressed up to essentially infinite times---we have tried various extremely large values of the evolution time using our exponentiation routine for time evolution and have found that the gauge violation remains in a steady state indefinitely. Even though for extreme errors the compliant sequence of coefficients has to be computed with high precision, we have illustrated how experimentally relevant local gauge breaking due to unassisted matter tunnelling or gauge flipping can be robustly protected against even when the sequence of coefficients nontrivially departs from a perfectly compliant sequence. 

Moreover, we have demonstrated the protection in a digital circuit implemented in Cirq. Also in this case, we have found excellent gauge-invariance protection up to the largest simulated evolution times of $20/J$, and we have established the optimal protection strength for given Trotter step size. 

Our results lend for a number of immediate extensions. 
They can be applied to any Abelian lattice gauge theory and to higher powers of the gauge-symmetry generators. Thus, the same protective power holds for the conventionally proposed, but experimentally much more challenging, two-body protection scheme that is quadratic in the generators of Gauss's law, as well as for $\mathrm{Z}_2$ gauge theories. 
The method can also be immediately generalized to protect global symmetries. 
Moreover, we have related our results to DD and EGP, showing that these can enjoy a much stronger protective power against coherent errors than previously known. Conversely, DD and EGP for stabilizer codes have been shown to protect well against $1/f$ noise \cite{nakamura_charge_2002,Young2013}, which is ubiquitous in solid state systems~\cite{paladino_mathbsf1mathbsfitf_2014}.
Since the spectrum density of $1/f$ noise is mostly concentrated in the low-frequency range, the DD sequence does not need to be ultra-fast~\cite{shiokawa_dynamical_2004}; similarly, the energy gap needed to suppress $1/f$ is moderate. It has been demonstrated experimentally that DD can be used to improve gate fidelity~\cite{west_high_2010,zhang_protected_2014,pokharel_demonstration_2018}. Since our scheme can be interpreted as DD in the time-dependent case, it can be used in a similar manner to suppress the $1/f$ noise.
Finally, we have discussed how controlled gauge violation in a recent cold-atom experiment \cite{Yang2020} can be reinterpreted in the light of our method, yielding an elegant interpretation of gauge protection in that experiment as well as clear guidelines on how to improve it in future works. 
With its experimental simplicity and high flexibility, and having a firm theoretical framework behind it, the proposed single-body gauge protection thus shows a clear way forward to achieving controlled gauge invariance in modern gauge quantum simulators. 
As part of an ongoing study \cite{Halimeh2020d}, we expect the protection discussed here to present a localization transition, similar to many-body localization \cite{Abanin_review} and energy localization \cite{DAlessio2013,heyl2019quantum}.

In the current era of noisy intermediate-scale quantum devices, where fully scalable, universal, and fault-tolerant quantum computers are still out of reach, further progress hinges crucially on the design of error-mitigation strategies that can be implemented in existing hardware. In our work, we have designed such a strategy, which may enable quantum computers to study such complicated issues as the out-of-equilibrium dynamics of strongly-coupled gauge theories or the emergence of gauge invariance in nature.

\section*{Acknowledgements}The authors thank Bing Yang for helpful discussions, and acknowledge support by Provincia Autonoma di Trento, the DFG Collaborative Research Centre SFB 1225 (ISOQUANT), and the ERC Starting Grant StrEnQTh (Project-ID 804305). 

\appendix

\section{More about $\mathrm{U}(1)$ gauge theory in quantum link formalism}\label{sec:appMoreAboutU1QLM}
Possible eigenvalues of the Gauss's-law generators of the $\mathrm{U}(1)$ QLM given in Eq.~\eqref{eq:Gj} are $2, 1, 0, -1$ for every matter site $j$ up to a factor of $(-1)^j$. However, not all eigenvalue combinations are physically allowed. Up to a factor of $(-1)^j$, a local constraint at matter site $j$ with gauge-generator eigenvalue $2$ requires the matter site $j$ and field links $j-1,j$ and $j,j+1$ to be spin up, which forbids the gauge-generator eigenvalue $-1$ for its two neighbors. Hence, up to a factor of $(-1)^j$ there are no ``$2,-1$'' or ``$-1,2$'' combinations in any of the allowed gauge sectors.

	\section{Quantum Zeno effect}\label{sec:QZE}
	
	In this Appendix, we discuss how weaker but still well-controlled protection can be achieved even when relaxing the stringent requirement in Eq.~\eqref{eq:c_j} of the theorem discussed in Sec.~\ref{sec:theorem} of the main text. In particular, we present a formal framework based on the quantum Zeno effect for coherent systems, evolved in continuous time as well as Trotterized schemes.
	
	A sufficiently large $V$ restricts the system dynamics to the decoherence-free subspace of $H_G$, a phenomenon known as a continuous formulation of the quantum Zeno effect (QZE) \cite{facchi2002quantum,burgarth2019generalized}. More precisely, considering the Hamiltonian in  Eq.~\eqref{equ:effectiveHamiltonian}, we obtain 
	\begin{equation}
	\lim_{V\rightarrow\infty} e^{-itH} = e^{-it[VH_G + \sum_{n} \Pi_n(H_0 + \lambda H_1)\Pi_n]},
	\label{eqn:QZE}
	\end{equation}
	with a residual additive term of $\mathcal{O}(J^2L^2t/V)$. Here, $H_G$ need not necessarily have an integer spectrum as is required in Sec.~\ref{sec:theorem}, and it can encode any desired global symmetry. Now, we specialize to the protection of a target subspace of a local gauge symmetry.
	
	There are two situations where the QZE can promise protected dynamics up to a timescale $t\propto V/(JL)^2$, with a controlled violation of $\mathcal{O}(J^2L^2/V)$.  In the first situation, the spectrum of $H_G=\sum_j c_j G_j$ is nondegenerate. Specifically, for a general $H_1$, the dynamics is protected when the $c_j$ are sufficiently incommensurate, i.e., for arbitrary $\textbf{g}_1 \neq \textbf{g}_2$, $\textbf{c}^\intercal\cdot(\textbf{g}_1-\textbf{g}_2)\neq 0$ (here, we defined $\textbf{c}$ as the vector of $c_j$, as in the main text). This condition can be easily satisfied when $c_j$ are random numbers or irrational numbers, or even fine-tuned integers. 
	
In the second situation, $H_1$ cannot split up the degeneracy of the spectrum of $H_G$ at first-order perturbation theory. In this case, one has 
	$\Pi_nH_1\Pi_n =\sum_{\textbf{g},\textbf{g}\prime\in\{\textbf{g},\textbf{g}\prime|\textbf{c}^\intercal\cdot\textbf{g}=n,\textbf{c}^\intercal\cdot\textbf{g}\prime=n\}} P_\textbf{g}H_1P_{\textbf{g}\prime} = \sum_{\textbf{g}\in\{\textbf{g}|\textbf{c}^\intercal\cdot\textbf{g}=n\}} P_\textbf{g}H_1P_\textbf{g}$, 
	where we used the operators $\Pi_n = \sum_{\textbf{g}\in\{\textbf{g}|\textbf{c}^\intercal\cdot\textbf{g}=n\}} P_\textbf{g}$ that project on the subspaces of fixed eigenvalues $n$ of $H_G$, with $P_\textbf{g}$ the projector on gauge sector $\textbf{g}$.  
	When the above condition is satisfied, $U(t) \sim \exp\{-i\sum_\textbf{g} [nVP_\textbf{g} + P_\textbf{g}(H_0 + \lambda H_1)P_\textbf{g}]t\}$ with $n=\textbf{c}^\intercal\cdot\textbf{g}$. This situation can make the sequences of $c_j$ much simpler. For instance, the physical error term Eq.~\eqref{eq:expH1} in the $\mathrm{U}(1)$ gauge theory considered in Eq.~\eqref{eq:H0} 
	conserves the parity and only causes the gauge violation $\{+1,-1\}$ or $\{-1,+1\}$ in pairs of nearest-neighbor sites. It is straightforward to verify that the coupling due to $H_1$ at leading order cannot split the degeneracy for $H_G$ when choosing all $c_j=(-1)^j$. The protective effect in such a situation can be clearly seen in, e.g., Fig.~\ref{fig:AnalogDynamics_experiment}(c) of the main text.
	 
	We can extend these considerations to digital quantum simulators. Digital quantum simulation with a protection term can be regarded as a quantum system undergoing ``kicks'' according to the evolution operator 
	\begin{equation}
	U_m(t) = [U_{\mathrm{kick}}U_0(t/m)]^m,
	\end{equation}
	where $U_0(t/m) = e^{-i(H_0+\lambda H_1)t/m}$, $U_{\mathrm{kick}} = e^{-iVH_Gt/m}$, and $t/m=\delta t$ is the Trotter time step. The spectrum decomposition of $U_{\mathrm{kick}}$ can be expressed as $U_{\mathrm{kick}}=\sum_n{e^{-inVt/m}\Pi_n}$, where a nondegeneracy condition $nVt/m\neq n' Vt/m \mod 2\pi$, $\forall n\neq n'$, is assumed. The unitary kicks version of the QZE states that in the large $m$ limit and for $V\sim \mathcal{O}(m/t)$, $U_m(t)\sim \exp\{-i\sum_n [nV\Pi_n + \Pi_n(H_0 + \lambda H_1)\Pi_n]t\}$ \cite{facchi2004unification,facchi2009quantum}. The evolution operator thus becomes identical to the evolution operator for the above ``continuous'' QZE. Hence, the protection sequences for analog quantum simulation also work for digital quantum simulation.
	In our numerics, we find that already a strong $V$ that is constant, i.e., not increasing with $m$, provides controlled protection over the simulated times.
	
	Notably, this protection due to the QZE effect is different from the slow rise of gauge invariance for the case when $\lambda$ is perturbatively \textit{small} as compared to $H_0$ \cite{Halimeh2020b,Halimeh2020c}. Indeed, $\lambda$ can be much stronger than the scales of $H_0$, as long as it is dominated by $V$. In this sense, the present case is an instance of \textit{strong perturbation theory}.
	
	\begin{figure}[!ht]
		\centering
		\includegraphics[width=.48\textwidth]{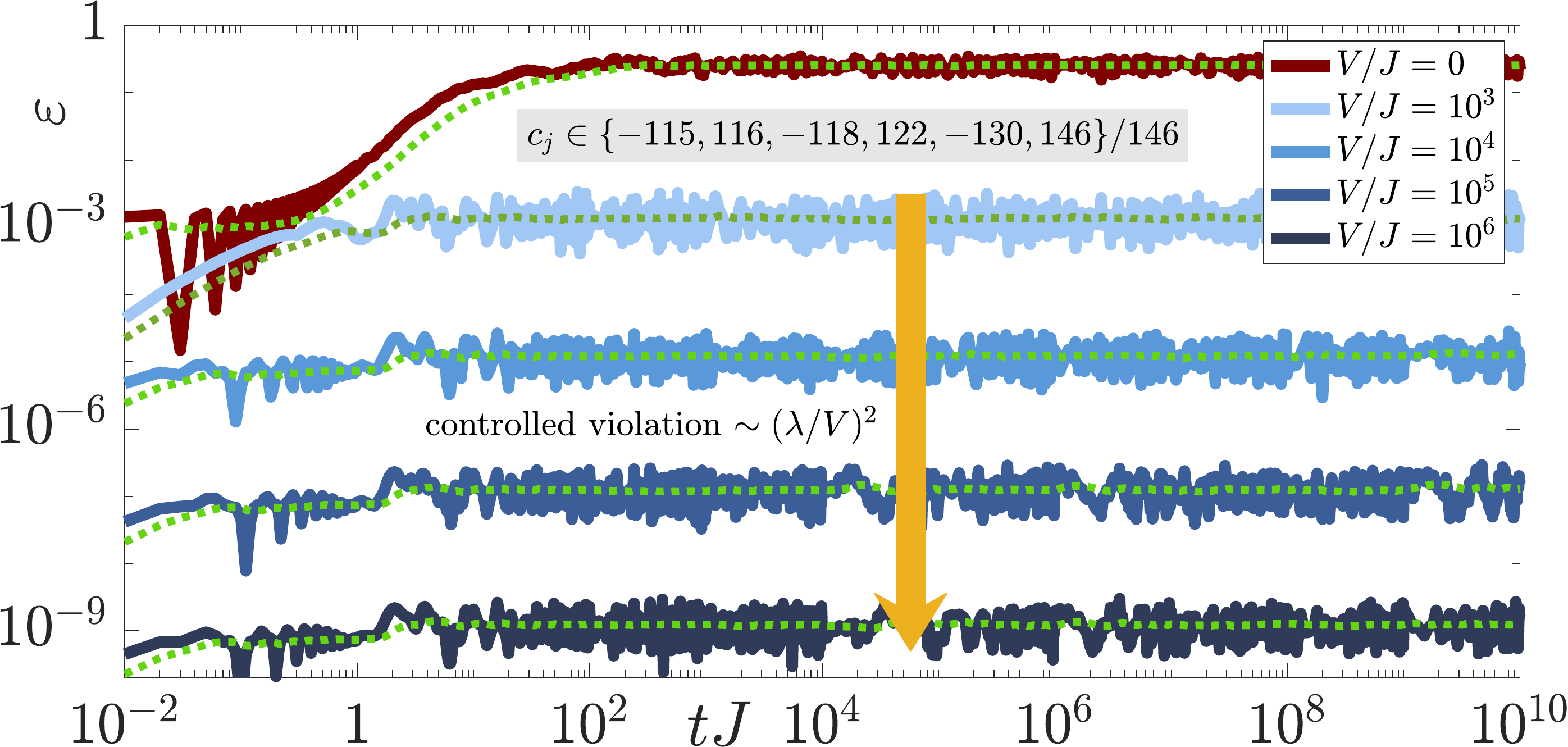}
		\caption{(Color online). Temporally nonaveraged gauge violations, whose temporal averages (marked here in dotted green lines) are shown in Fig.~\ref{fig:AnalogDynamics}(a). The qualitative picture is the same, especially since the finite-size fluctuations are rather small and suppressed with $V$ (note that our $y$-axis is on a log scale).}
		\label{fig:Nonaveraged}
	\end{figure}
	
		\begin{figure}[!ht]
		\centering
		\includegraphics[width=.48\textwidth]{{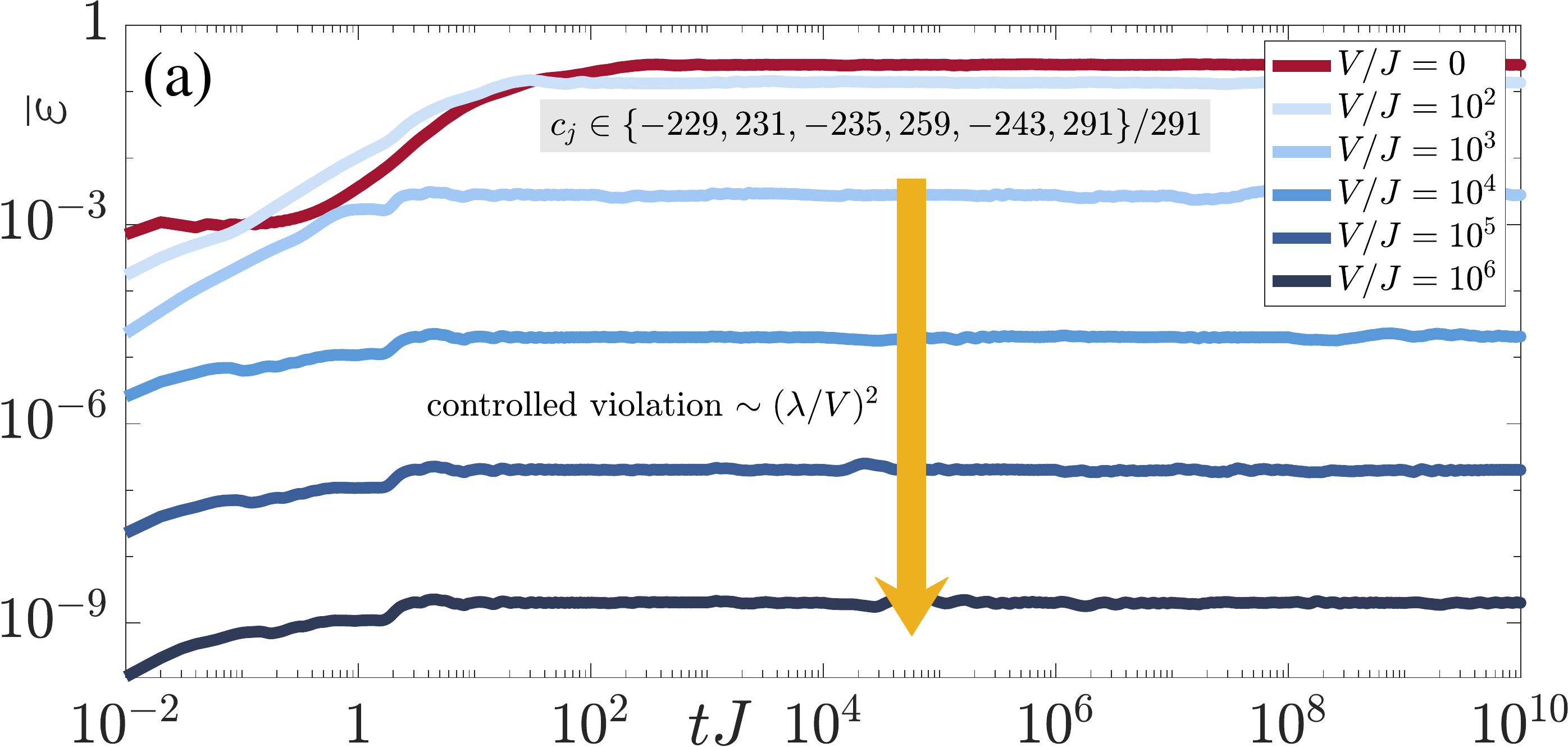}}\\
		\includegraphics[width=.48\textwidth]{{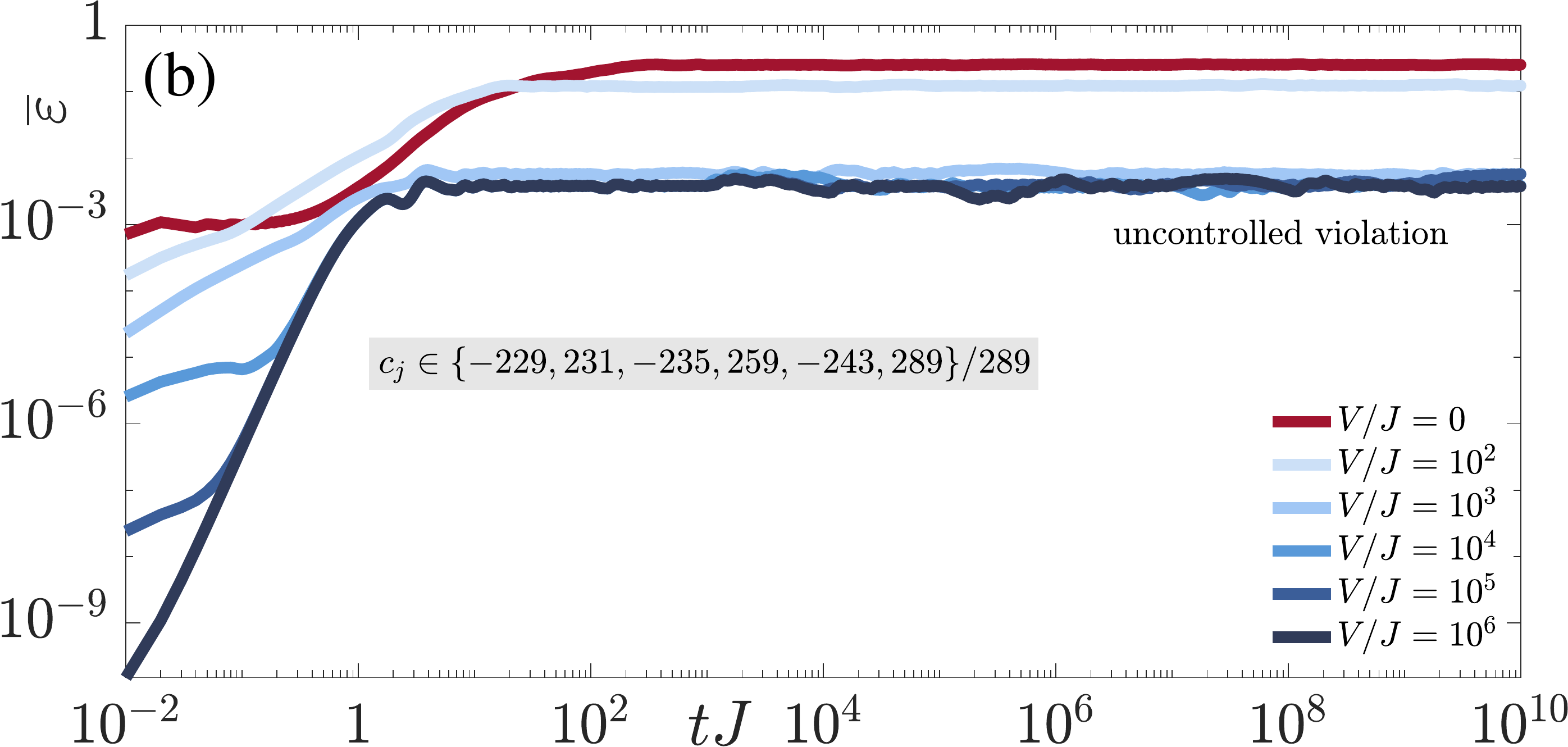}}\\
		\includegraphics[width=.48\textwidth]{{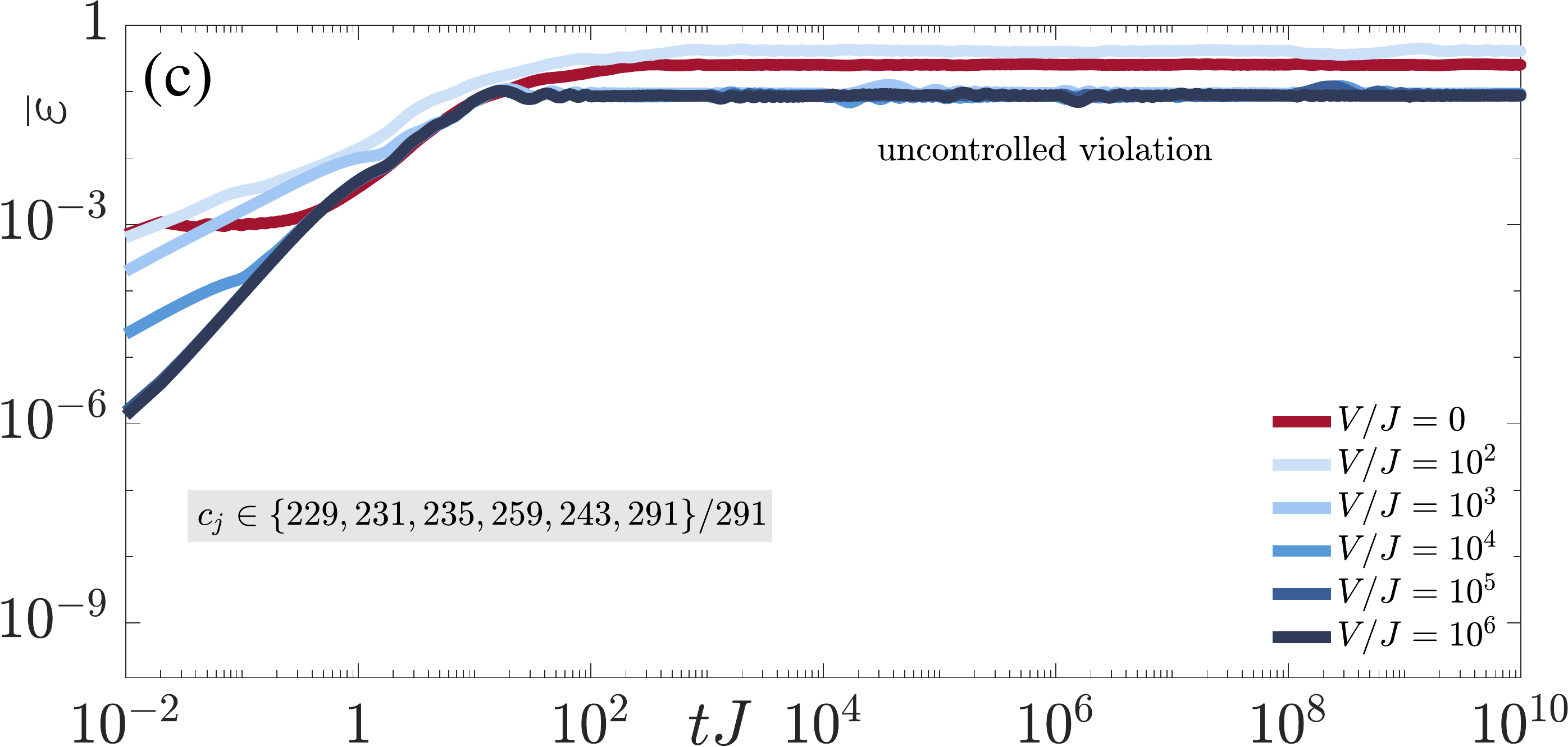}}
		\caption{(Color online). Same as Fig.~\ref{fig:AnalogDynamics} in the main text, but for (a) another compliant sequence, (b) another noncompliant sequence, and (c) a noncompliant sequence that is the nonstaggered version of the compliant sequence in (a). The results are qualitatively identical to those of Fig.~\ref{fig:AnalogDynamics}, with control over gauge invariance being achieved only for the compliant sequence in (a).}
		\label{fig:AnalogDynamicsSM}
	\end{figure}
	
	\begin{figure*}[!ht]
		\centering
		\includegraphics[width=.48\textwidth]{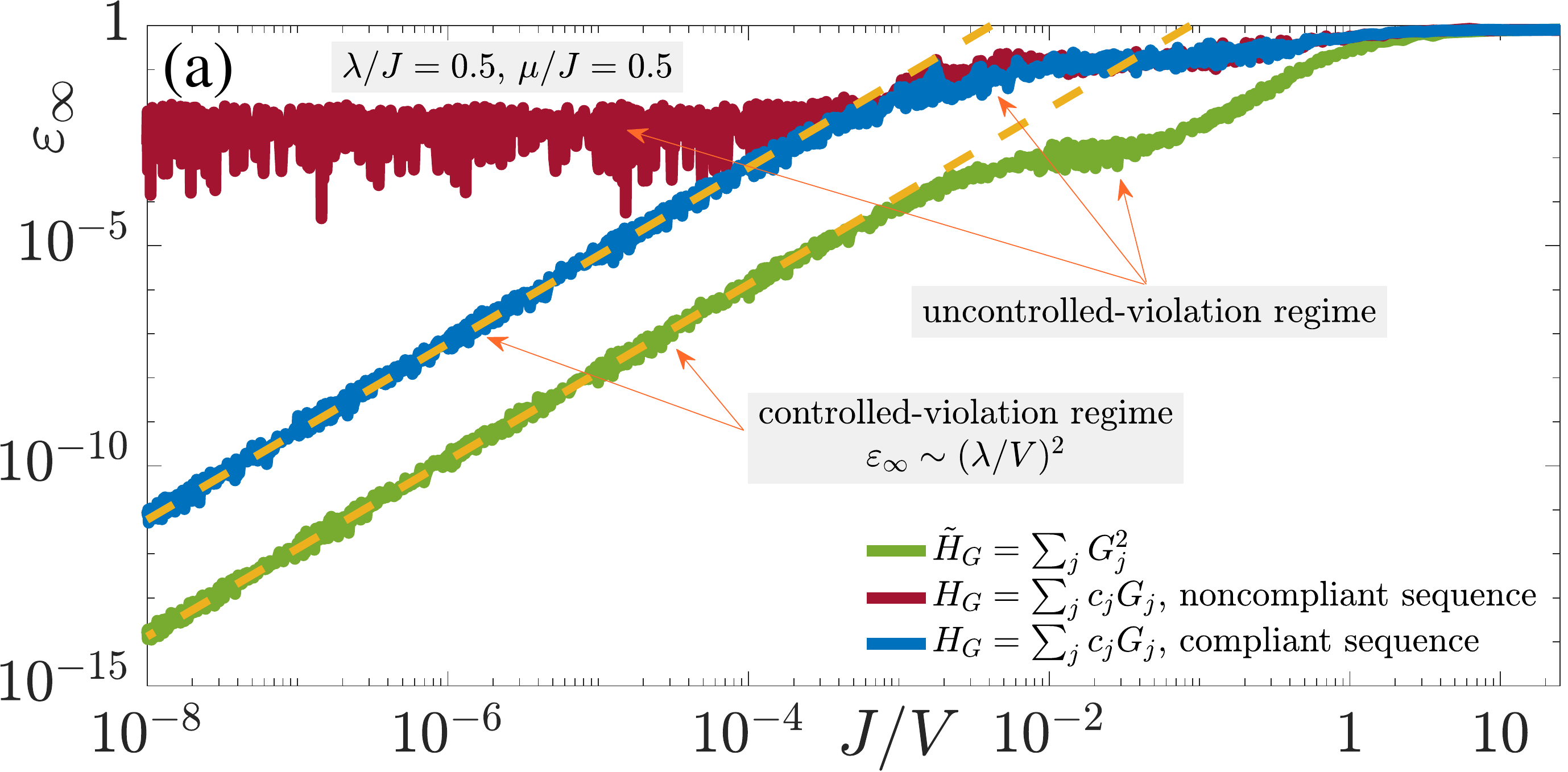}\quad
		\includegraphics[width=.48\textwidth]{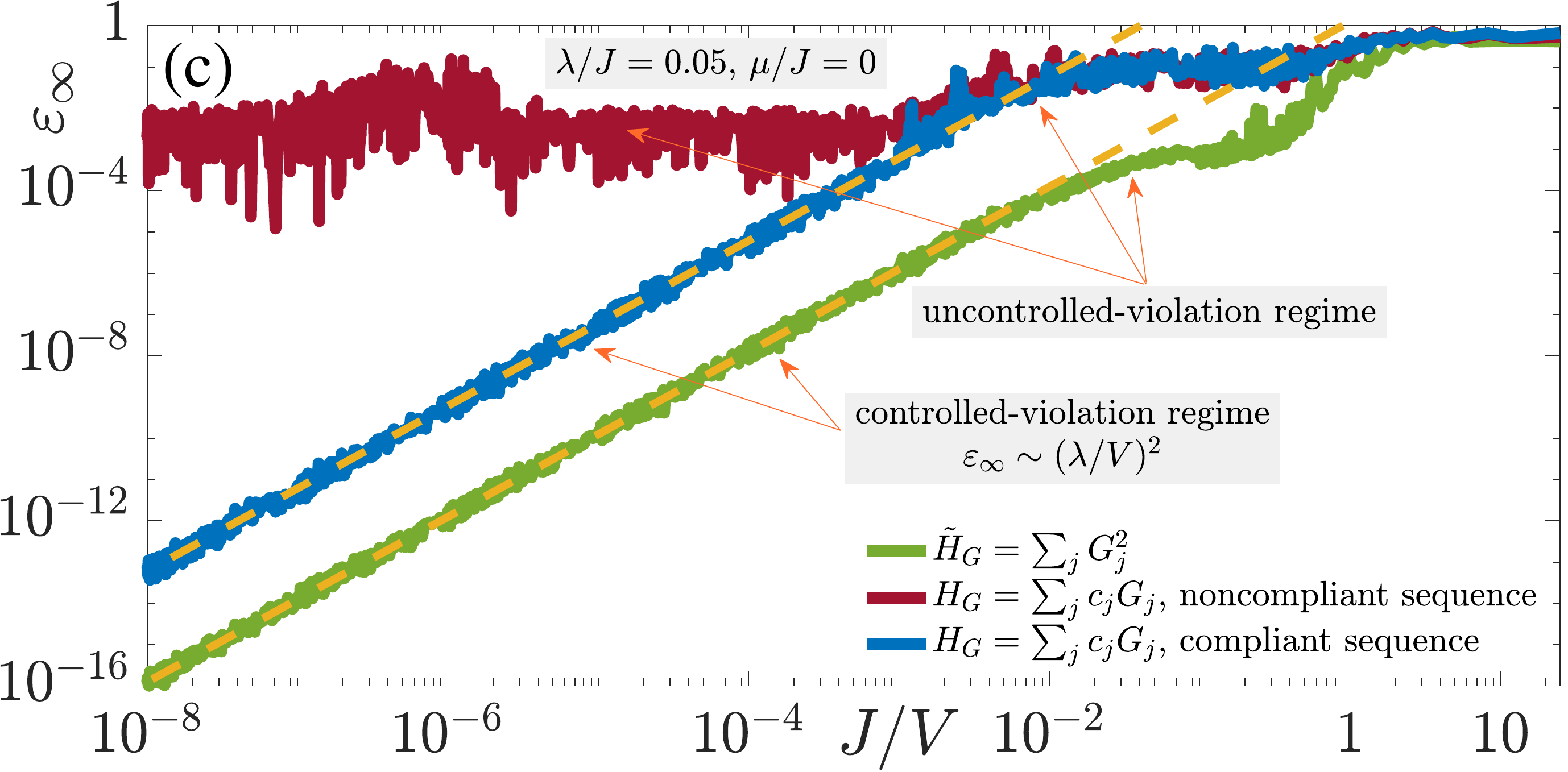}\quad\\
		\includegraphics[width=.48\textwidth]{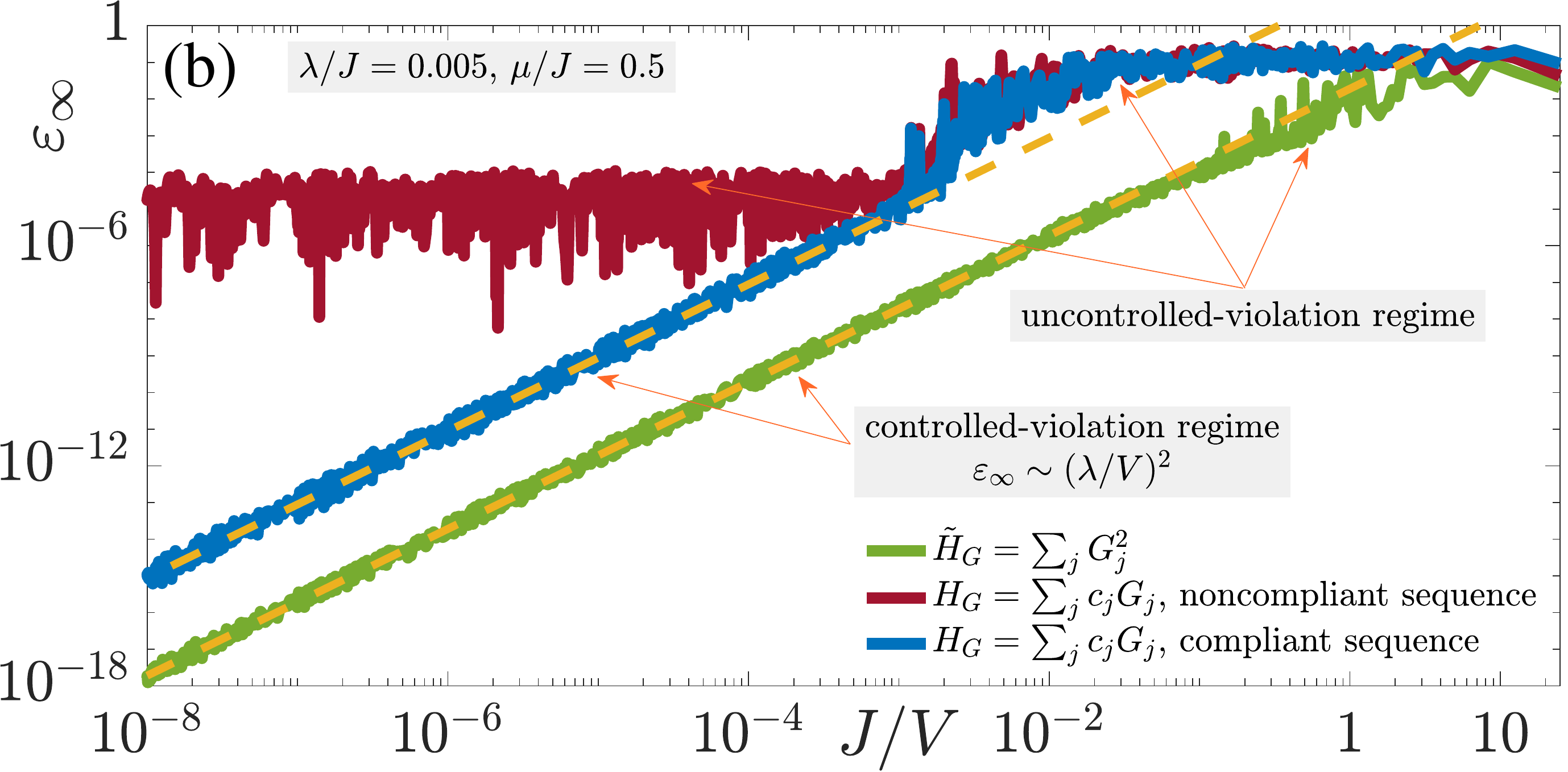}\quad
		\includegraphics[width=.48\textwidth]{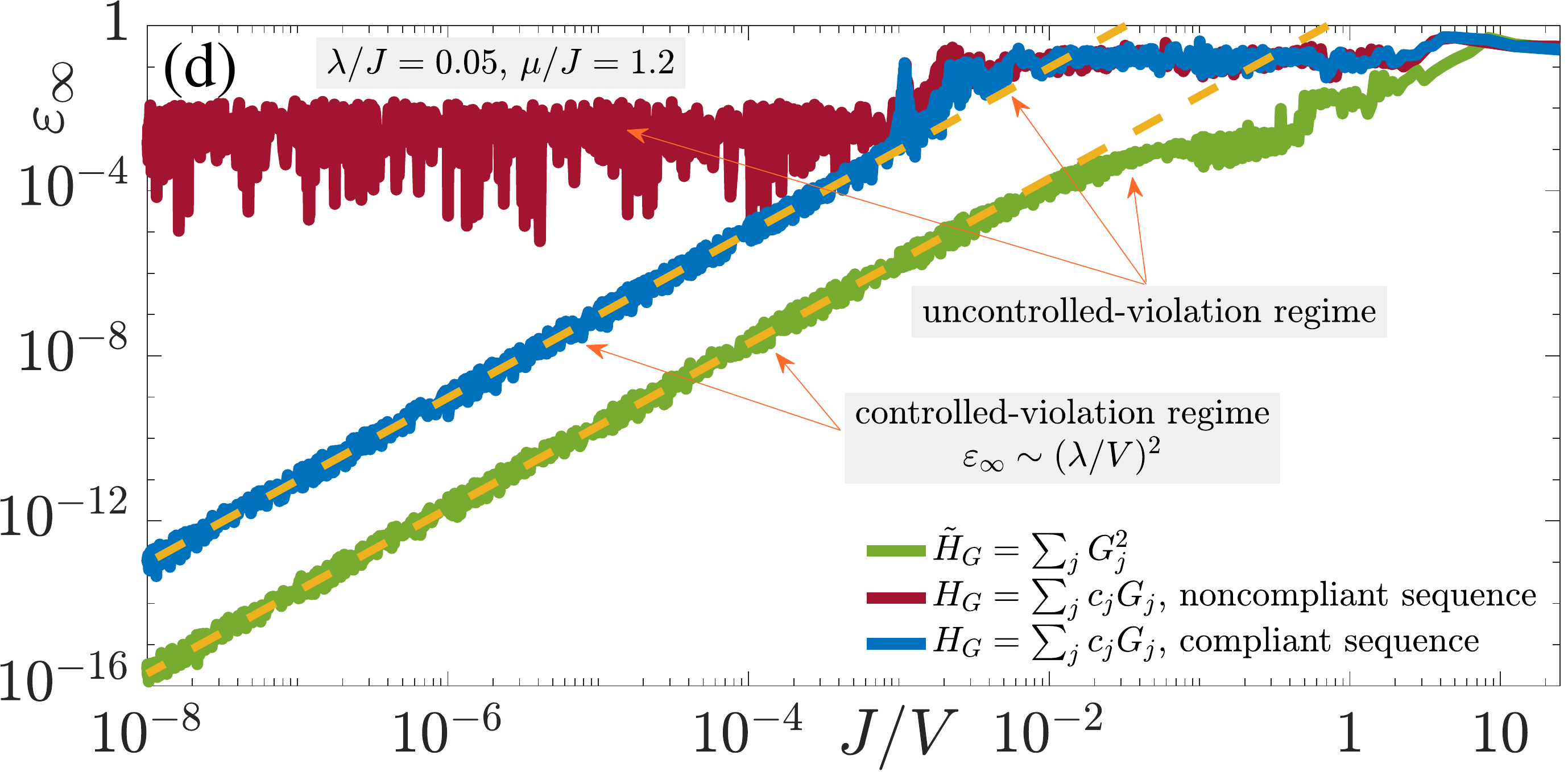}\quad
		\caption{(Color online). Same as Fig.~\ref{fig:AnalogScan} in the main text, but for (a) $\lambda/J=0.5$ and $\mu/J=0.5$, (b) $\lambda/J=0.005$ and $\mu/J=0.5$, (c) $\lambda/J=0.05$ and $\mu/J=0$, and (d) $\lambda/J=0.05$ and $\mu/J=1.2$. Comparing these results to those of Fig.~\ref{fig:AnalogScan}, the qualitative picture remains unchanged regardless of the values of $\lambda$ and $\mu$.}
		\label{fig:AnalogScanSM}
	\end{figure*}
	
	\begin{figure}[!ht]
		\centering
		\includegraphics[width=.48\textwidth]{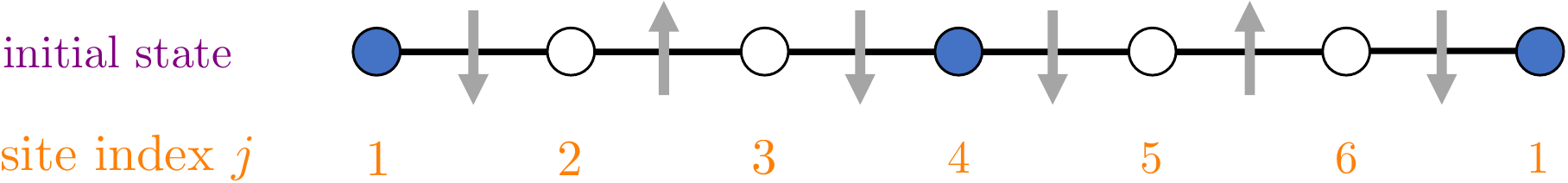}\\
		\includegraphics[width=.48\textwidth]{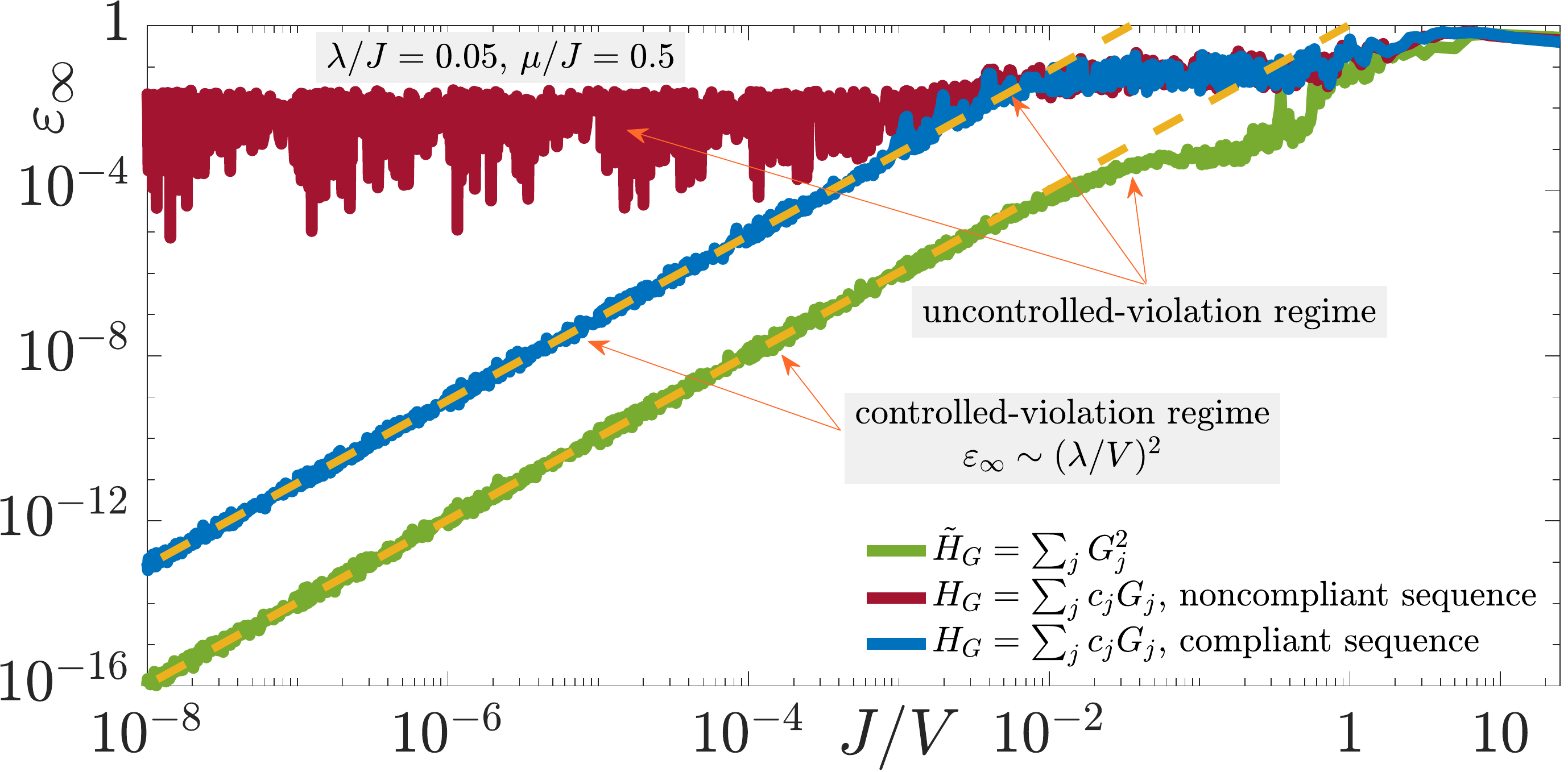}
		\caption{(Color online). Same as Fig.~\ref{fig:AnalogScan} but for the initial state drawn on top. The qualitative conclusion is identical to that of Fig.~\ref{fig:AnalogScan}, showing that our findings are independent of the gauge-invariant initial state.}
		\label{fig:OtherInitial}
	\end{figure}

\section{Numerics specifics}\label{sec:specs}
For benchmarking a potential analog quantum simulator, we have used the QuTiP \cite{Johansson2012,Johansson2013} exact diagonalization toolkit in order to construct the model and initial state, but for the time evolution we have opted for our own exponentiation routine that is better suited for handling the very large evolution times we access. Even though in the main text we show mostly results for the temporally averaged violation $\overline\varepsilon(t)=\int_0^t\d s\,\varepsilon(s)/t$, with $\varepsilon(s)$ given in Eq.~\eqref{eq:violation}, the temporally nonaveraged violation exhibits the same behavior as shown in Fig.~\ref{fig:Nonaveraged}, albeit in the presence of finite-size fluctuations, which are, however, suppressed with $V$.

In the case of the digital circuit, we make use of the quantum circuit library Cirq \cite{cirq}. We construct the circuit sketched in Fig.~\ref{fig:QuantumCircuit} and simulate its full wave function with readout of the gauge violation $\varepsilon$ occurring after each Trotter step.

\section{Further results on analog dynamics}\label{sec:AnalogSM}
In this Appendix, we corroborate the generality of our qualitative conclusions in the main text by showing results for different initial states and parameter values.

\subsection{Violation dynamics for different sequences of coefficients}
Here, we provide results for different compliant and noncompliant sequences than those used in the main text in the case of the ``extreme'' gauge-breaking error of Eq.~\eqref{eq:H1}. The corresponding results are shown in Fig.~\ref{fig:AnalogDynamicsSM}. Similarly to Fig.~\ref{fig:AnalogDynamics}(a), the gauge violation is controlled $\sim(\lambda/V)^2$ at large protection strength $V$ only when the sequence is compliant, i.e., it satisfies the condition $\sum_jc_jg_j=0\,\,\mathrm{iff}\,\,g_j=0,\,\forall j$, given in Eq.~\eqref{eq:c_j}, as shown in Fig.~\ref{fig:AnalogDynamicsSM}(a). Minor variations to this sequence will completely compromise this control of the violation, as shown in Fig.~\ref{fig:AnalogDynamicsSM}(b). Again, if the staggering is removed from the compliant sequence the associated violation is not controlled; see Fig.~\ref{fig:AnalogDynamicsSM}(c).

\subsection{Violation scan for different values of $\lambda$ and $\mu$}\label{sec:AnalogSM_scan}
In the main text, we have set $\lambda=0.05J$ and $\mu=0.5J$. In Fig.~\ref{fig:AnalogScan} we have shown the ``infinite-time" violation $\varepsilon_\infty$ as a function of inverse protection strength $J/V$ in the presence of ``extreme'' gauge breaking given in Eq.~\eqref{eq:H1}, under two-body and single-body gauge protection. Our conclusions hold for other values of the microscopic parameters $\lambda$ and $\mu$, as shown in Fig.~\ref{fig:AnalogScanSM}, where we use the same compliant and noncompliant sequences $c_j\in\{-115,116,-118,122,-130,146\}/146$ and $c_j\in\{-115,116,-118,130,-122,145\}/145$, respectively, in the single-body protection. For sufficiently large $V$, the gauge violation is controlled $\sim(\lambda/V)^2$ in the case of two-body as well as compliant-sequence single-body gauge protection. The single-body protection with the noncompliant sequence cannot bring the dynamics perturbatively close to the ideal gauge theory, but rather seems to bring about a lower bound in $\varepsilon_\infty$ regardless of how large $V$ is.

\subsection{Violation scan for a different initial state}
In the main text, we have focused on the initial state shown on top of Fig.~\ref{fig:AnalogDynamics}, which comprises empty matter sites with the gauge links pointing along the positive or negative $z$-direction in a staggered fashion. Here, we repeat the results of Fig.~\ref{fig:AnalogScan} for a different initial state containing particles on matter sites $j=1,4$, with the links between these two sites carrying the configuration $\downarrow\uparrow\downarrow$ (as throughout the paper, periodic boundary conditions are assumed); see top of Fig.~\ref{fig:OtherInitial}. The corresponding ``infinite-time'' violations as a function of $J/V$ for the extreme error of Eq.~\eqref{eq:H1} are shown in Fig.~\ref{fig:OtherInitial}. The two-body protection and its single-body counterpart with the compliant sequence $c_j\in\{-115,116,-118,122,-130,146\}/146$ give rise to a controlled-violation regime for sufficiently large $V$, where $\varepsilon_\infty\sim(\lambda/V)^2$, bringing the model perturbatively close to a renormalized version of the ideal gauge theory described by $H_0$ of Eq.~\eqref{eq:H0}. Single-body protection with the noncompliant sequence does not provide control over the violation regardless of how large $V$ is.

\subsection{Violation scan for experimentally relevant local errors}\label{sec:AnalogSM_scan_expError}

\begin{figure}[!ht]
	\centering
	\includegraphics[width=.48\textwidth]{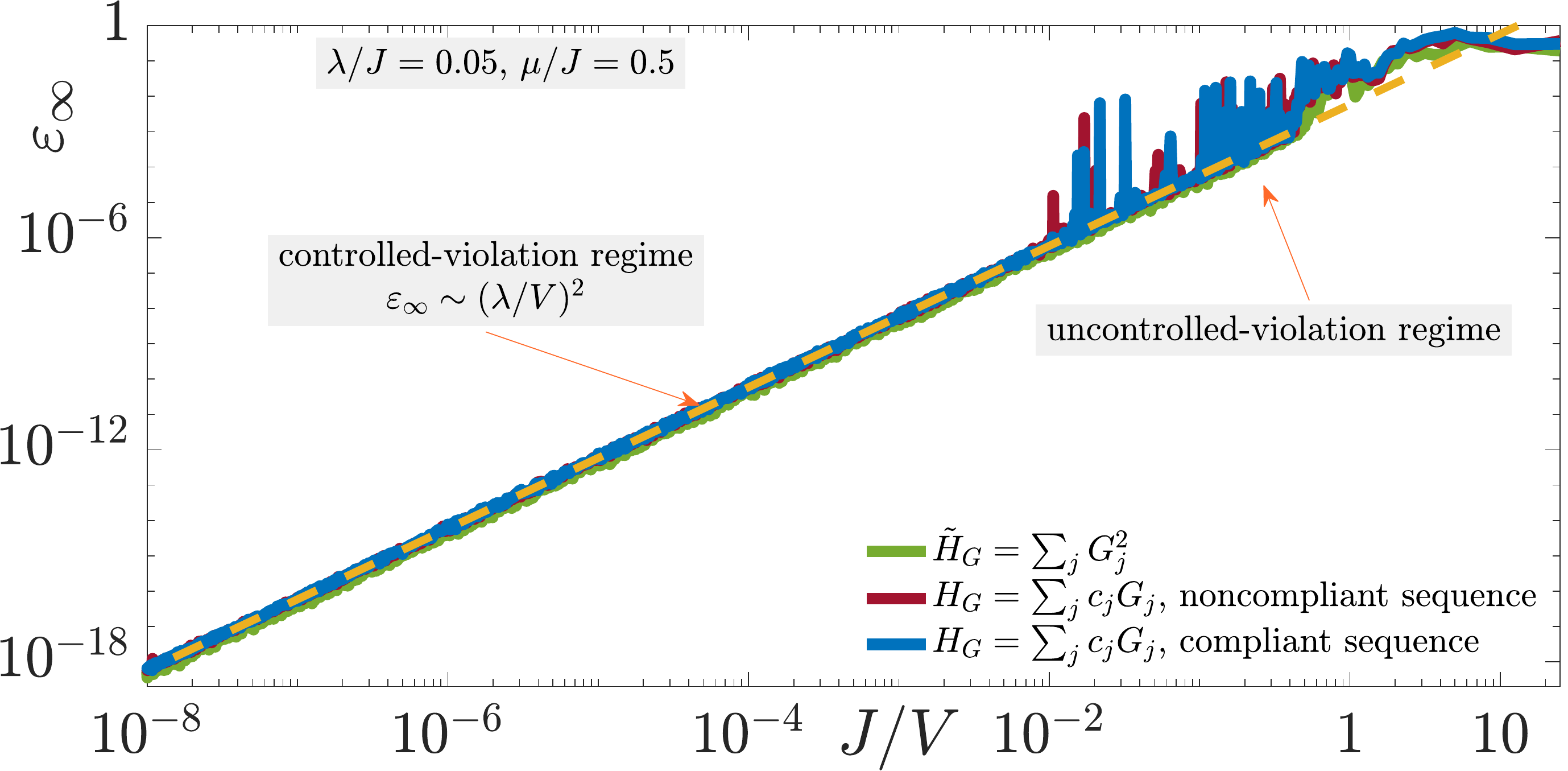}
	\caption{(Color online). Same as Fig.~\ref{fig:AnalogScan} but for the experimentally relevant local gauge-breaking term $H_1=\sum_j(\tau^x_{j,j+1}+\sigma^+_j\sigma^+_{j+1}+\sigma^-_j\sigma^-_{j+1})$. Even though the noncompliant sequence fails to achieve a controlled gauge violation in the case of the nonlocal gauge-breaking term of Eq.~\eqref{eq:H1} (see Fig.~\ref{fig:AnalogScan}), here it performs as well as the compliant sequence. As another marked difference from Fig.~\ref{fig:AnalogScan}, the linear protection performs as well as the full protection in the case of the local gauge-breaking term in Eq.~\eqref{eq:expH1}.}
	\label{fig:AnalogScan_experiment}
\end{figure}

\begin{figure}[!ht]
\centering
\includegraphics[width=.48\textwidth]{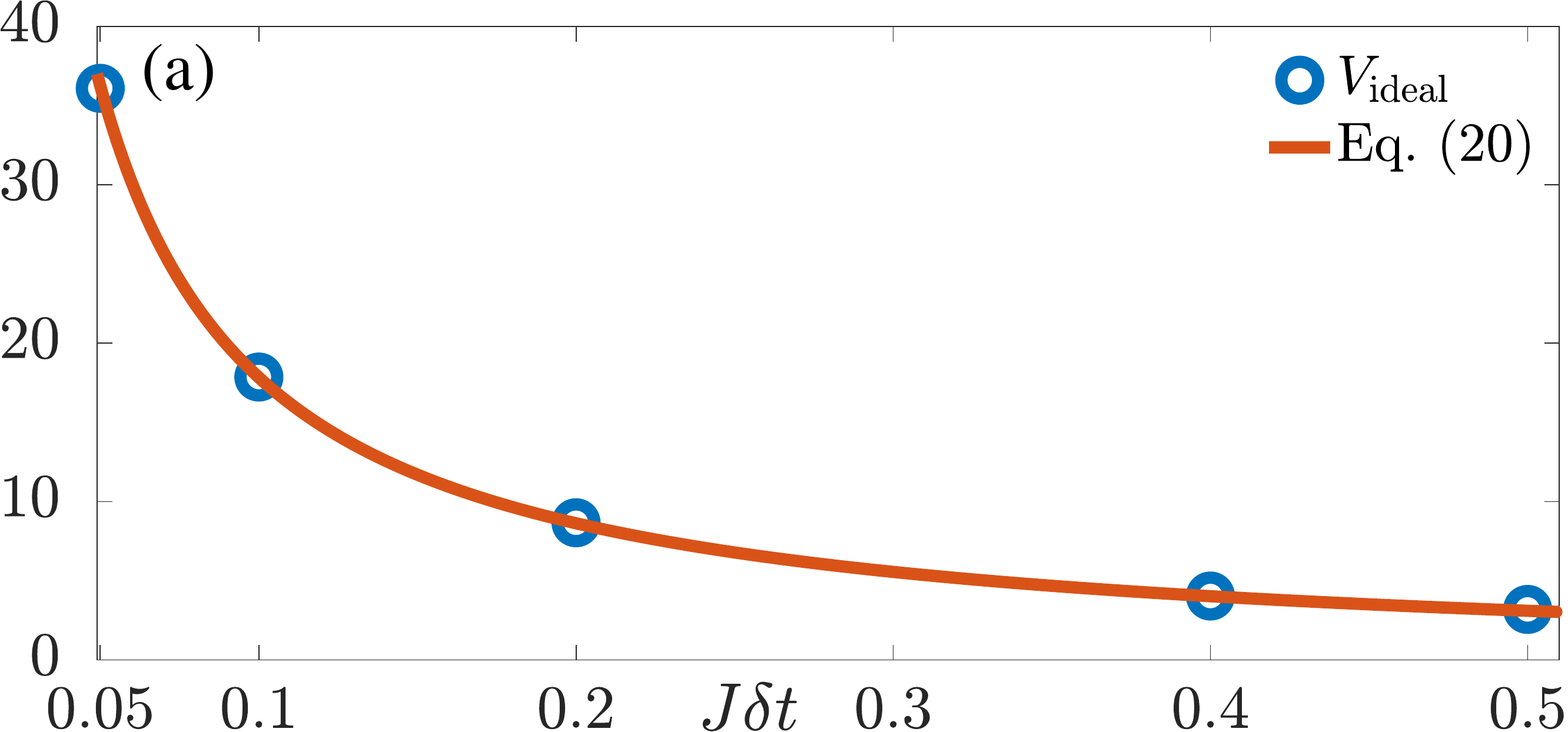}\\
\includegraphics[width=.48\textwidth]{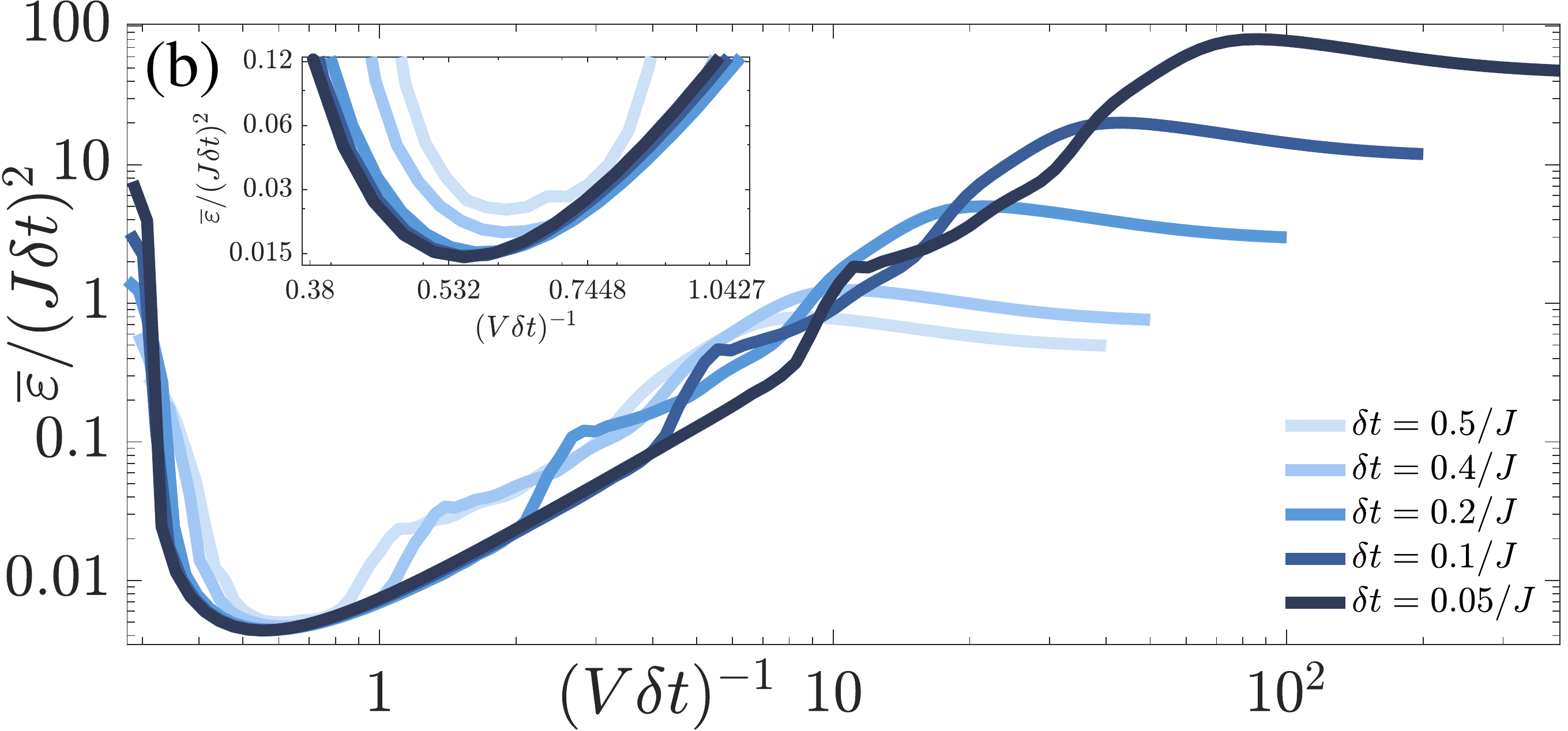}
\caption{(Color online). (a) Ideal protection strength $V_{\mathrm{ideal}}$ that provides the minimal temporally averaged gauge violation $\overline{\varepsilon}$. Blue diamonds are numerically extracted data points for the gauge-violation Hamiltonian of Eq.~\eqref{eq:expH1} and $\lambda=0.05J$, $\mu=0.5J$. The red line is given by Eq.~\eqref{eq:Videal}, $V_\mathrm{ideal}\approx\pi/(2\overline{c}\delta t)-\xi$, where the offset depending on microscopic parameters, $\xi$, has been determined by a fit as $\approx0.58J$. (b) Rescaled mean gauge violation depending on protection strength for various Trotter time steps $\delta t$. The results collapse around their minimum.
}
\label{fig:Videal}
\end{figure}

In the main text and Appendix~\ref{sec:AnalogSM_scan}, we have shown results for the ``infinite-time'' gauge violation as a function of $J/V$ in the case of an extreme nonlocal gauge-breaking term given in Eq.~\eqref{eq:H1} protected against by using either the two-body or single-body energy penalty given in Eqs.~\eqref{eq:FullProtection} and~\eqref{eq:LinearProtection}, respectively. However, as discussed in the main text, experimentally relevant errors are usually milder than Eq.~\eqref{eq:H1} and are dominated by local terms, such as those of Eq.~\eqref{eq:expH1}. It is expected that the two-body energy penalty $V\tilde{H}_G=V\sum_jG_j^2$ and the single-body protection term $VH_G=V\sum_jc_jG_j$ with a compliant sequence $\{c_j\}$---i.e., $\sum_jc_jG_j\ket{\psi}=0,\,\,\mathrm{iff}\,\,G_j\ket{\psi}=0,\,\forall j$---will still lead to a controlled-violation regime for sufficiently large protection strength $V$, given that the gauge-breaking error in Eq.~\eqref{eq:expH1} is much more forgiving than its counterpart in Eq.~\eqref{eq:H1}. This is indeed the case, as shown in Fig.~\ref{fig:AnalogScan_experiment}.

Nevertheless, a fundamental difference arises in the case of the experimentally relevant local error of Eq.~\eqref{eq:expH1} with respect to the extreme gauge breaking of Eq.~\eqref{eq:H1}: now even the noncompliant sequence can offer controlled violation for sufficiently large $V$. This is a promising finding for experimental purposes, as it means that there is room for imprecision in implementing the coefficients $c_j$, and that the condition $\sum_jc_jG_j\ket{\psi}=0,\,\,\mathrm{iff}\,\,G_j\ket{\psi}=0,\,\forall j$ is only a sufficient but not necessary condition in the case of experimentally relevant local errors.

\section{Ideal protection strength for digital circuit}
\label{sec:VidealSM}

As described in Sec.~\ref{sec:digitalCircuit} on the digital circuit implementation, the periodic degeneracy of gates with respect to their angle gives rise to a finite ideal protection strength $V_\mathrm{ideal}$. We find $V_\mathrm{ideal}$ to be given by Eq.~\eqref{eq:Videal}, which consists of a term $\propto\delta t^{-1}$ and a nonuniversal offset $\xi$ that depends on $\mu$ and the microscopic details of the gauge-breaking term $H_1$. 

As displayed in Fig.~\ref{fig:Videal}(a), Eq.~\eqref{eq:Videal} accurately reproduces the numerically extracted $V_\mathrm{ideal}$ for a wide range of Trotter time steps $\delta t$. Here, we use the exact same Hamiltonian and parameters as in Sec.~\ref{sec:digitalCircuit}, and we determine $\xi$, the single open parameter of Eq.~\eqref{eq:Videal}, by a fit. (The resulting $\xi$ is close to $\mu$.) 

The results depicted in Fig.~\ref{fig:Digital}(b) display a universal behavior when rescaling the mean gauge violation as $\overline{\varepsilon}\rightarrow\overline{\varepsilon}/(J\delta t)^2$ and the protection strength as $V\rightarrow V\delta t$. Under this rescaling, one observes a collapse of the gauge violation for all $\delta t$ around their minimum at $V_\mathrm{ideal}\delta t$; see Fig.~\ref{fig:Videal}(b). The universal behavior comes about due to the dominance of the first term of Eq.~\eqref{eq:Videal} (together with the rescaling of the gauge violation as $\varepsilon\sim\lambda^2/V^2$), while the nonuniversal additive constant $\xi$ provides only a comparativley small offset.

\bibliography{LinearProtectionBiblio,additional_references}
\end{document}